\documentclass[english,aps,prl,twocolumn,superscriptaddress,showpacs,amsmath]{revtex4-2}
\usepackage[T1]{fontenc}
\usepackage{amssymb}
\usepackage{bm}
\usepackage{amsmath}
\usepackage{mathtools}
\usepackage{graphicx}
\usepackage{babel}
\usepackage{xcolor}
\usepackage{units}
\usepackage{esint}
\usepackage{natbib}
\usepackage[breaklinks=true,colorlinks=true,urlcolor=blue,
citecolor=blue,linkcolor=blue,bookmarks=false]{hyperref}
\usepackage{lineno}

\setcitestyle{journalcolor = blue}
\date{18 August 2025}   
\begin{document}

\title{Dynamically tunable hydrodynamic transport in boron nitride-encapsulated graphene}

\author{Akash Gugnani} \thanks{equal contribution}
\email{akashgugnani@iisc.ac.in}
\affiliation{Department of Physics, Indian Institute of Science, Bangalore 560012, India}
\author{Aniket Majumdar} \thanks{equal contribution}
\email{aniketm@iisc.ac.in}
\affiliation{Department of Physics, Indian Institute of Science, Bangalore 560012, India}
\author{Kenji Watanabe}
\affiliation{Research Center for Electronic and Optical Materials, National Institute for Materials Science, 1-1 Namiki, Tsukuba 305-0044, Japan}
\author{Takashi Taniguchi}
\affiliation{Research Center for Materials Nanoarchitectonics, National Institute for Materials Science, 1-1 Namiki, Tsukuba 305-0044, Japan}
\author{Arindam Ghosh}
\email{arindam@iisc.ac.in}
\affiliation{Department of Physics, Indian Institute of Science, Bangalore 560012, India}
\affiliation{Center for Nano Science and Engineering, Indian Institute of Science, Bangalore 560012, India}

\begin{abstract}
    Over the past decade, graphene has emerged as a promising candidate for exploring the viscous nature of electronic flow facilitated by the availability of extremely high-quality devices employing a graphene channel encapsulated within dielectric layers of hexagonal boron nitride (hBN). However, the level of disorder in such systems is mainly determined by the device fabrication protocols, making it impossible to obtain a tunability between the impurity-dominated and the viscous transport within the same device. In this work, using a combination of ultraviolet (UV) radiation and gate electric field, we have demonstrated a dynamic modulation of charge hydrodynamics, quantified in the thermal and electrical transport by the extent of departure from the Wiedemann-Franz (WF) Law in monolayer graphene devices at room temperature. We achieved this by tuning the disorder level continuously and reversibly using UV light to create transient trap states in the encapsulating hBN dielectric. With progressive UV radiation, we observed a dramatic increase in the momentum-relaxing scattering relative to that between the electrons and also the Lorentz number, by nearly a factor of ten, with increasing disorder, thereby approaching the restoration of the WF law in highly disordered graphene. Our experiments outline a potent strategy to tune the fundamental mechanism of charge flow in state-of-the-art graphene devices.
\end{abstract}

\maketitle

\section{Introduction}

The hydrodynamic flow of electrons in mesoscopic systems has garnered significant attention in the past decade, owing to the emergence of ultra-clean two-dimensional materials like graphene \cite{dean2010boron,lin2019towards}, transition metal dichalcogenides \cite{wang2012electronics,wang2022making}, etc. The fluidic motion of electrons, characterised by strong momentum-conserving electron-electron interactions, has been detected experimentally through the observation of Planckian scattering \cite{gallagher2019quantum}, parabolic velocity profile \cite{ku2020imaging, vool2021imaging, sulpizio2019visualizing}, negative local resistance \cite{bandurin2016negative}, giant thermal diffusivity \cite{block2021observation}, Hall viscosity \cite{li2022hydrodynamic, berdyugin2019measuring}, WF law violation \cite{crossno2016observation}, electron vortices \cite{palm2024observation}, universal viscosity-to-entropy density lower bound \cite{majumdar2025universality}, etc. A common principle advocated by these experiments for realising electron hydrodynamics is that the electron-electron scattering timescale ($\tau_\mathrm{ee}$) must be much shorter than any momentum-relaxing scattering process ($\tau_\mathrm{mr}$), like electron-impurity scattering, diffuse scattering from the boundary, electron-phonon scattering etc. However, recent experiments \cite{aharon2022direct} have reported viscous electron flow even while violating the above criterion, opening up alternative routes towards electron hydrodynamics, such as para-hydrodynamics \cite{estrada2024alternativeq}, retroreflected hole backflow \cite{PhysRevB.109.085126}, and tomographic electron transport \cite{hofmann2022collective, zeng2024quantitative}. Therefore, it is imperative to study the interplay between momentum-conserving and momentum-relaxing collisions to understand and engineer the electron hydrodynamics in graphene.

\begin{figure*}[tbh]
    \centering
    \includegraphics[width=1.0\linewidth]{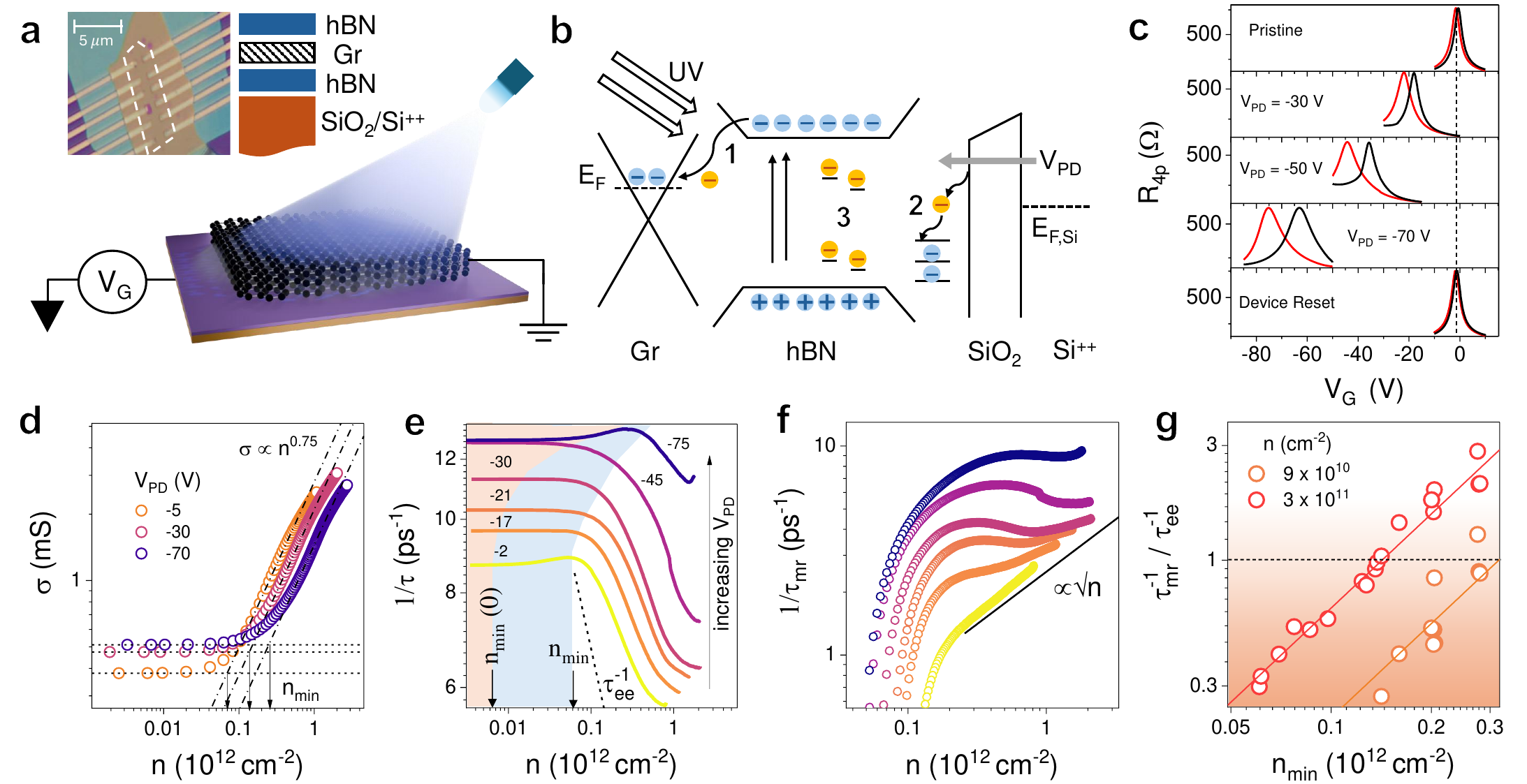}
    \caption{\textbf{Tuning electronic disorder using UV radiation:} (a). Top-left corner shows the optical micrograph of one of the measured devices (D3). The cross-sectional view of the heterostructure is shown to its right. [Below] Schematic of the experimental setup used for performing UV radiation-based tuning of disorder. (b) Different mechanisms responsible for charge redistribution upon UV illumination (indicated using thick white arrows with black border). Process 1 refers to UV-induced photogating, which results in the transfer of charges from hBN to graphene or vice versa. Process 2 indicates the formation of charge puddles in the mid-gap states of hBN as a result of UV excitation. Process 3 marks the onset of UV-induced activation of trap states at the hBN-SiO$_2$ interface and subsequent random trapping of charges at those sites. The blue-coloured charges are intrinsic charges residing in the different layers, while the yellow-coloured ones are the UV radiation-induced charges which have been redistributed across the heterostructure. (c) Variation of four-terminal resistance $R_\mathrm{4p}$ as a function of gate voltage $V_\mathrm{G}$ for both forward and backward $V_\mathrm{G}$ sweeps at four different values of $V_\mathrm{PD}$. The panel at the bottom shows the characteristics obtained by resetting the device to pristine condition (indicated by a vertical dotted line) by shining UV radiation at $V_\mathrm{PD} = 0$~V. (d) Variation of electrical conductivity ($\sigma$) with $n$ in device D3 for different values of $V_\mathrm{PD}$. The resulting $n_\mathrm{min}$ for each of these traces has been obtained by noting the intersection between the lines $\sigma = \sigma_\mathrm{min}$ (dotted line) and $\sigma \propto n^{0.75}$ (dashed line), where $\sigma_\mathrm{min}$ is the electrical conductivity at the Dirac point. (e) Variation of the total scattering rate $\tau^{-1}$ as a function of $n$ for different values of $V_\mathrm{PD}$. The dashed line shows the variation of electron-electron relaxation scattering rate $\tau_\mathrm{ee}^{-1}$ with $n$. The $n<n_\mathrm{min}(0)$ region is colored with light red, whereas $n_\mathrm{min} (0) \le n \le n_\mathrm{min}$ region is colored with light blue. (f) Variation of the momentum relaxation scattering rate $\tau_\mathrm{mr}^{-1}$ as a function of $n$ for the same values of $V_\mathrm{PD}$, as those depicted in panel (e). The solid line indicates a $\sqrt{n}$ dependence. (g) Ratio of momentum relaxation scattering rate ($\tau_\mathrm{mr}^{-1}$) to the electron-electron scattering rate ($\tau_\mathrm{ee}^{-1}$) as a function of $n_\mathrm{min}$ for two different carrier densities. The solid lines vary as $n_\mathrm{min}$ and serve as a guide to the eye. The colored region corresponding to $\tau_\mathrm{mr}^{-1}/\tau_\mathrm{ee}^{-1} < 1$ indicates the viscous limit, while the region with $\tau_\mathrm{mr}^{-1}/\tau_\mathrm{ee}^{-1} > 1$ indicates the diffusive limit.}
    \label{fig1}
\end{figure*}

Modern device fabrication techniques \cite{wang2013one, wang2022mechanical, lin2011clean} have made it possible to significantly minimise the intrinsic disorder concentration and charge inhomogeneity (indicated by $n_\mathrm{min}(0)$) in exfoliated two-dimensional layers. For monolayer graphene, $n_\mathrm{min}(0)$ can go down to as low as $10^9$~cm$^{-2}$ \cite{xin2023giant, nam2017electron} in state-of-the-art field effect transistor (FET) devices. However, existing graphene FETs suffer from a lack of tunability of the impurity concentration as the $n_\mathrm{min}(0)$ is solely determined by the fabrication protocols. Recent efforts to explore the controlled implantation of disorder in bad metals utilising helium ion irradiation \cite{jin2020disorder} were successful, but they lead to permanent degradation of the sample, thereby impeding the demonstration of a reversible hydrodynamic-to-diffusive crossover. UV radiation, on the other hand, has been used as a tuning knob \cite{neumann2016spatial, ju2014photoinduced, luo2012controlled} to transfer charges across the different layers of graphene-hBN-based van der Waals (vdW) heterostructures. These transferred charges can operate as free carriers and contribute to the electrical current, or get trapped within the dielectric, causing fluctuations in the background potential, leading to the formation of charge inhomogeneity across the sample. Moreover, by applying a gate electric field \cite{roy2013graphene, quezada2020persistent}, one can control the number of carriers getting transferred as well as the direction of transfer, thereby creating a new avenue for achieving precisely reversible disorder-doping interplay of the electronic channel. 

In this work, we have chosen high-quality monolayer graphene FETs encapsulated within multi-layered hBN flakes as the platform for realising electron hydrodynamics. Further, we have employed UV radiation as a tool to controllably modulate the extent of charge disorder subjected to the graphene layer and have explored the mechanisms of the electrical transport at room temperature. Using noise thermometry, we have found that the introduction of disorder tends to restore the thermal conductivity to the classical Wiedemann-Franz (WF) \cite{ashcroft1978solid} value, given by $\kappa_e = \mathcal{L}_\mathrm{WF} \,\sigma T$ where $\kappa_e$ is the electronic contribution to thermal conductivity, $\sigma$ is the electrical conductivity and $\mathcal{L}_\mathrm{WF} = \pi^2 k_\mathrm{B}^2/3e^2$, with $k_\mathrm{B}$ and $e$ being the Boltzmann constant and the electronic charge respectively. This technique allows one to reversibly tune a high-quality graphene device from the hydrodynamic regime into the diffusive regime and back, at a temperature as high as $300$~K, providing crucial insights on the interplay of momentum-conserving and momentum-relaxing collisions in the emergence of viscous electronic flow.

\section{Electrical characterisation of UV-irradiated graphene layers}

Mechanically exfoliated graphene flakes, after encapsulation using hBN, were transferred onto a Si$^{++}$/SiO$_2$ substrate and etched into rectangles ($\simeq 20$~$\mu$m $\times 10$~$\mu$m), and electrical contact was established using edge-contacted gold electrodes (Fig.~\ref{fig1}a, see also Supplementary Section S1). The total charge inhomogeneity present in these vdW heterostructures at temperature $T$ is given by $n_\mathrm{min}$ where $n_\mathrm{min} = n_\mathrm{min}(0) + n_\mathrm{th}(T)$, with $n_\mathrm{th}(T)=(2\pi^3/3)(k_\mathrm{B}T/hv_\mathrm{F})^2$ referring to the thermally excited carrier density at temperature $T$ \cite{xin2023giant, ponomarenko2024extreme, majumdar2025universality} ($h$, $k_\mathrm{B}$ and $v_\mathrm{F}$ are the Planck's constant, Boltzmann's constant and Fermi velocity respectively). The high quality of the devices can be attributed to the low intrinsic charge inhomogeneity values, quantified using the residual charge carrier density $n_\mathrm{min}(0) \simeq 7 \times 10^{9}$~cm$^{-2}$. Additionally, the strain-induced gap in the density of states was estimated to be too small to impact our experiments, which were carried out at room temperature. (See Supplementary Section S2). To tune the background potential fluctuations, the graphene devices were subjected to a gate electric voltage (called the photo-doping voltage $V_\mathrm{PD}$) and uniformly irradiated using UV radiation (Fig.~\ref{fig1}a) with wavelength $\lambda = 235$~nm and a constant power density $\approx 1$~pW/~$\mu$m$^2$ (Details in Supplementary Section S3). The consequent light-matter interaction within the encapsulating hBN layers results in the redistribution of charges within the vdW heterostructure via two different mechanisms (Fig.~\ref{fig1}b) - 

(1) \textbf{Interlayer charge transfer} - The band gap ($E_\mathrm{g}$) of hBN is $\simeq 5$~eV \cite{cassabois2016hexagonal}. Upon UV illumination ($hc/\lambda \gtrsim E_\mathrm{g}$), electron-hole pairs are generated and the applied electric field prevents the recombination of photo-induced carriers, leading to a migration of electrons (holes) from hBN to graphene (indicated in Fig.~\ref{fig1}b as Process-$1$), under the application of a negative (positive) $V_\mathrm{PD}$. This process is identical to photogating \cite{roy2013graphene}; as a result, the Dirac point shifts close to $V_\mathrm{PD}$ for both forward and backwards gate voltage ($V_\mathrm{G}$) sweeps (See Supplementary Section S4-a). Upon applying a $V_\mathrm{PD}$ as large as $-70$~V to the Si substrate, the graphene channel undergoes photo-doping by nearly $3 \times 10^{11}$~cm$^{-2}$ - a process generating $\approx 10^9$~cm$^{-2}$ carriers per unit volt of $V_\mathrm{PD}$, which corresponds to a shift in Fermi energy ($E_\mathrm{F}$) by just $6.5$~meV.(See Supplementary Section S4-b).

(2) \textbf{Random charge trapping} - UV radiation can also activate random shallow and deep trap states at the hBN-SiO$_2$ interface \cite{ahmed2020generic} (indicated in Fig.~\ref{fig1}b as Process-$2$) alongside localised charge puddles within the hBN layer \cite{uddin2017probing, du2015origin, du2016origins} (shown in Fig.~\ref{fig1}b as Process-$3$), resulting in a decrease in carrier mobility ($\mu$) of the channel (See Supplementary Section S4-c and S4-d) by almost an order of magnitude - from $2$~m$^2$~V$^{-1}$~s$^{-1}$ to $0.3$~m$^2$~V$^{-1}$~s$^{-1}$ as $V_\mathrm{PD}$ is swept from $0$~V to $-70$~V. Additionally, in both forward and backwards gate voltage sweeps, the transfer characteristics show an increasing anti-hysteresis and transform from a sharp to a broader Lorentzian peak, with increasing $V_\mathrm{PD}$ (Fig.~\ref{fig1}c). This increase in full width at half maximum (FWHM) of the peak is directly proportional to the increasing total charge inhomogeneity ($n_\mathrm{min}$), which has been evaluated from the $x$-intercept of the intersection between $\sigma = \sigma_\mathrm{min}$ and $\sigma \propto n^{\gamma}$ ($\gamma \simeq1$), where $\sigma_\mathrm{min}$ is the electrical conductivity at the Dirac point (Fig.~\ref{fig1}d).

UV radiation can also lead to permanent degradation of the graphene layer \cite{gao2014defect}. However, we observe that the graphene channel with its CNP around $-70$~V (Fig.~\ref{fig1}c), under UV illumination with $V_\mathrm{PD} = 0$~V, experiences a shift in Dirac point from $\approx -75$~V to $\approx -0.5$~V and the pristine electrical characteristics of the graphene FET are revived (Fig.~\ref{fig1}c, bottom-most panel). This restoration of the disorder level to the pristine condition via UV-induced extrinsic hole doping (Details in Supplementary Section S5) confirms the reversible nature of disorder tunability of this technique.

\begin{figure*}[tbh]
    \centering
    \includegraphics[width=1.0\linewidth]{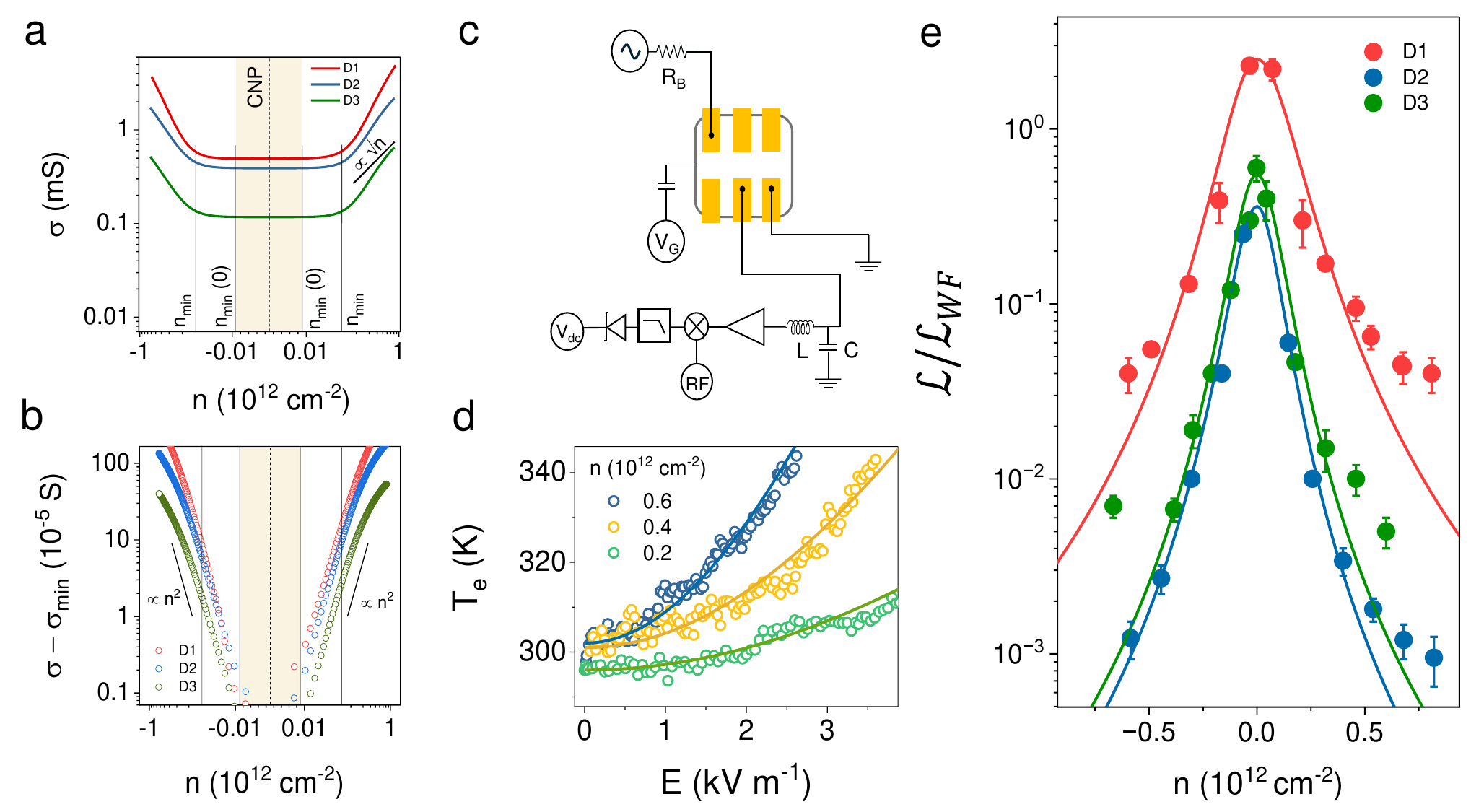}
    \caption{\textbf{WF law violation as a signature of electron hydrodynamics:} (a) Variation of $\sigma$ as a function of $n$ for three different devices D1, D2 and D3. The solid line (colored in black) indicates a $\sqrt{n}$ dependence and serves as a guide to the eye. $n_\mathrm{min}$ and $n_\mathrm{min}(0)$ have been indicated using vertical solid lines. The area shaded in light pink indicates $n < n_\mathrm{min}(0)$. (b) Variation of $\sigma - \sigma_\mathrm{min}$ as a function of $n$ for three traces shown in (a). The solid line (colored in black) indicates a $n^2$ dependence and serves as a guide to the eye. The area shaded in light pink indicates $n < n_\mathrm{min}(0)$. (c) Schematic of the electrical circuit used for Johnson noise thermometry measurements. (d) Variation of $T_\mathrm{e}$ as a function of $E$ for D2, at three different carrier concentrations given by $n (10^{12}$~cm$^{-2}) = 0.2, 0.4, 0.6$. (e) Variation of $\mathcal{L}/\mathcal{L}_\mathrm{WF}$ as a function of $n$ for three devices, D1, D2 and D3, all in pristine condition. The solid curves refer to the Lorentzian fit of the data, as per Eqn.~\ref{L}. The error bars associated with the data points represent the standard deviation of results obtained from the experiment.}
    \label{fig2}
\end{figure*}

The combined effect of the above-mentioned mechanisms leads to a competition between the momentum relaxation scattering rate ($\tau^{-1}_\mathrm{mr}$) of the electrons in the disordered graphene device and the momentum conserving electron-electron scattering rate ($\tau^{-1}_\mathrm{ee}$), which is determined only by $T$ when $T > T_\mathrm{F}$ (and $T_\mathrm{F}$ when $T_\mathrm{F} > T$). $\tau^{-1}_\mathrm{mr}$ is obtained from the experimental quasiparticle relaxation rate $\tau^{-1}$ (Fig.~\ref{fig1}e), which is given by $\tau = \sigma \hbar k_\mathrm{F}/v_\mathrm{F} ne^2$, with $n$, $\hbar$, $k_\mathrm{F}$ and $e$ being the net carrier density measured relative to the instantaneous charge neutrality point (CNP), the reduced Planck's constant, Fermi wave vector and the electronic charge respectively. Applying Matthiessen's rule, in the leading order, we get $\tau^{-1} = \tau^{-1}_{mr} + \tau^{-1}_{ee}$ where $\tau^{-1}_{ee}$ is the electron-electron scattering rate, which, in the presence of a finite Fermi surface, is given by $\tau^{-1}_\mathrm{ee} \approx (\alpha k_\mathrm{B}T/\hbar v_\mathrm{F})^2 (v_\mathrm{F}/k_\mathrm{F})$ ($\alpha \approx 0.5$ \cite{xie2016transport, ghahari2016enhanced} being the fine structure constant). The resulting $\tau^{-1}_{mr}$ is plotted in Fig.~\ref{fig1}f as a function of $n$ for the same set of $V_\mathrm{PD}$ used in Fig.~\ref{fig1}e. In pristine graphene, momentum relaxation at high electron densities is driven by electron-acoustic phonon scattering \cite{kaasbjerg2012unraveling, perebeinos2010inelastic}, and the scattering rate scales as 
\begin{equation} \label{eq:electron phonon scattering}
\tau^{-1}_\mathrm{el-ph}=\frac{1}{\hbar^2}\frac{E_\mathrm{F}}{4 v_\mathrm{F}}\frac{D^2}{\rho_\mathrm{m} v_\mathrm{ph}^2} k_\mathrm{B}T \propto \sqrt{n}
\end{equation}
where $\rho_\mathrm{m}$ is the mass density of graphene, $v_\mathrm{ph}$ is the acoustic phonon velocity, $E_\mathrm{F}$ is Fermi Energy, and $D$ is the deformation potential \cite{hwang2008acoustic}. The experimental data points, as indicated in Fig.~\ref{fig1}f, agree well with Eqn.~\ref{eq:electron phonon scattering} and the acoustic phonon deformation potential $D$ is evaluated to be $\simeq 15$~eV, which agrees with previous experimental estimates \cite{bolotin2008temperature, efetov2010controlling}. The addition of disorder to the channel provides one more pathway to relax the momentum - Coulomb force-mediated electron-impurity scattering \cite{dassarma2011electronic, sarkar2015role, gosling2021universal}, yielding $\tau^{-1}_\mathrm{el-imp} \propto \sqrt{n}$ (Details in Supplementary Section S6). Finally, the ratio of the momentum-relaxing and momentum-conserving scattering rates, denoted by $\tau_\mathrm{mr}^{-1}/\tau_\mathrm{ee}^{-1}$ in Fig.~\ref{fig1}g, as a function of UV-induced disorder $n_\mathrm{min}$ demonstrates the crossover of electrical transport in the graphene channel from the clean limit ($\tau_\mathrm{mr}^{-1}/\tau_\mathrm{ee}^{-1} < 1$) to the dirty limit ($\tau_\mathrm{mr}^{-1}/\tau_\mathrm{ee}^{-1} > 1$). We note that the critical disorder density required to achieve this crossover depends on the Fermi energy level of the graphene channel, with doped graphene ($n \simeq10^{11}$~cm$^{-2}$) requiring lower amounts of disorder while undoped graphene ($n \approx 10^{10}$~cm$^{-2}$) experiencing momentum-conserved electronic scattering over the entire range of experimentally accessible disorder concentrations. 

\section{Dependence of the Electrical and Thermal conductivity on disorder}

\begin{figure*}[tbh]
    \centering
    \includegraphics[width=1.0\linewidth]{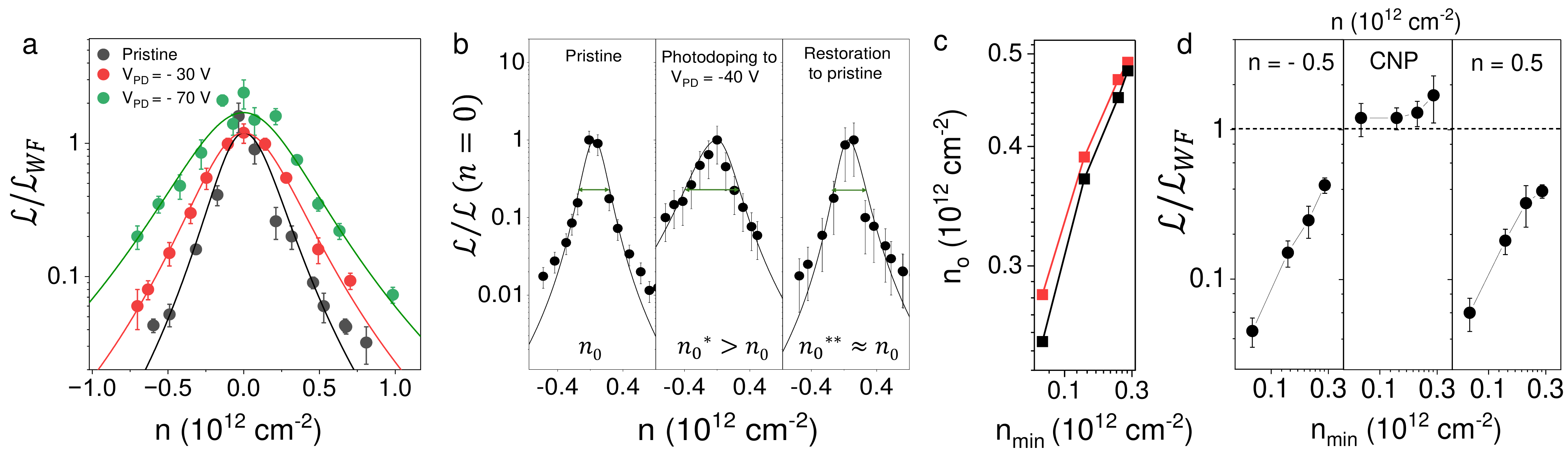}
    \caption{\textbf{Signature of dynamically-tunable electron hydrodynamics in thermal transport:} (a) Variation of $\mathcal{L}/\mathcal{L}_\mathrm{WF}$ as a function of n for device D3 at different values of $V_\mathrm{PD}$. The solid curves refer to the Lorentzian fits of the data, as per Eqn.~\ref{L}. The error bars associated with the data points represent the standard deviation of results obtained from the experiment. (b) $\mathcal{L}/\mathcal{L} (n=0)$, where $\mathcal{L}(n=0)$ refers to the value of $\mathcal{L}$ at the CNP, as a function of $n$ for three different scenarios: (Left) Pristine graphene, (Center) After the pristine graphene has been UV-irradiated with $V_\mathrm{PD} = -40$~V, (Right) After the UV-irradiated graphene is again subjected to UV radiation with $V_\mathrm{PD} = 0$~V, thereby restoring the pristine conditions. The FWHMs ($n_0, n_0^*, n_0^{**}$) of the Lorentzian fits and the comparison of their magnitudes are mentioned near the bottom of each panel.  The error bars presented here indicate the standard deviation of the experimental data points. (c) Variation of $n_0$ as a function of $n_\mathrm{min}$ for device D3. (d) Variation of $\mathcal{L}/\mathcal{L}_\mathrm{WF}$ as a function of $n_\mathrm{min}$ for device D3, part of which has been shown in (a). The three panels refer to three distinct carrier density regimes: hole-doped (Left), CNP (Center) and electron-doped (Right). The dashed line refers to $\mathcal{L}/\mathcal{L}_\mathrm{WF} = 1$.  The error bars presented reflect the standard deviation of the experimental data points.}
    \label{fig3}
\end{figure*}

We first explore the nature of charge and heat transport in the pristine hBN/Gr/hBN FETs, prior to the irradiation with UV. In all the measured devices, $\sigma$ as a function of $n$ at $300$~K, shown in Fig.~\ref{fig2}a, has similar features - (a) $n \le n_\mathrm{min}(0)$: the inhomogeneous or disordered regime, where $\sigma = \sigma_\mathrm{min}$. (b) $n_\mathrm{min}(0) \le n \le n_\mathrm{min}$: the thermal regime, where thermally excited electrons and holes govern the number of carriers dictating $\sigma$. In Fig.~\ref{fig2}b, we show the $n$-dependence of $\sigma-\sigma_\mathrm{min}$, where the thermal regime is characterised by $\sigma-\sigma_\mathrm{min} \propto n^2$, expected for the quantum critical hydrodynamic transport where the momentum-conserving scattering dominates the relaxation of current \cite{muller2009graphene, majumdar2025universality}. (c) $n \ge n_\mathrm{min}$: the Fermi liquid regime where $\sigma$ is controlled by free carriers generated by the gate voltage $V_\mathrm{G}$. Though the channel tends towards ballisticity, with $\sigma \propto \sqrt{n}$ at high densities, there is a small window at the intermediate carrier densities where $\sigma-\sigma_\mathrm{min} \propto n^2$, suggestive of a mono-component viscous fluid. In this regime, at low temperatures, we also observe quadratic dependence of $\sigma$ on $W$ (width of the channel) and negative local resistance, further establishing the viscous nature of electron transport \cite{majumdar2025universality}. These characteristics, however, usually get suppressed at temperatures as high as $300$~K. 

We subsequently measured the electronic thermal conductivity in these high-quality graphene devices with Johnson noise thermometry \cite{majumdar2025universality, crossno2016observation}, a well-established primary thermometry technique for accurate estimation of the electronic temperature ($T_\mathrm{e}$) in graphene. The schematic of the noise thermometry circuit is shown in Fig.~\ref{fig2}c (See Supplementary Section S7 for details). At certain values of $V_\mathrm{G}$, we have performed Joule heating across the electronic channel by applying an in-plane DC electric field ($E$) and measured $T_\mathrm{e}$ as a function of $E$ (Fig.~\ref{fig2}c). Solving the heat diffusion equation \cite{majumdar2025universality} for this geometry, we get
\begin{equation}
    T_\mathrm{e} = \dfrac{2\cal L}{3L^2E^2}\left[\left(T_\mathrm{c}^2 + \dfrac{L^2}{\cal L} E^2 \right)^{3/2} - \left(T_\mathrm{c}^2\right)^{3/2}\right]
    \label{Te}
\end{equation}
where $L$, $\cal L$ and $T_\mathrm{c}$ are respectively the length of the channel, effective Lorentz number and electron temperature at the metallic contact. By fitting $T_\mathrm{e}$ against $E$ using Eqn.~\ref{Te}, we estimate the $\mathcal{L}$ at different densities for the electron fluid in graphene. The plot of $\mathcal{L}/\mathcal{L}_\mathrm{WF}$ as a function of $n$ for three different devices (Fig.~\ref{fig2}e) is reproducible across all our high-quality samples in pristine condition and exhibits a strong violation of WF law. Since the charge and heat flow pathways are decoupled in the presence of strong momentum-conserving scattering, this violation is considered to be a signature of viscous electronic flow \cite{crossno2016observation, PhysRevLett.115.056603}. As per relativistic hydrodynamics, for a fluid of Dirac fermions \cite{lucas2018hydrodynamics,li2022hydrodynamic}, $\mathcal{L}/\mathcal{L}_\mathrm{WF}$ varies as 

\begin{equation}
    {\cal L}(n,T) = \frac{1}{e^2}{{\left [\frac{s(T)n_0(T)}{n^2+n^2_0(T)}\right ]}}^2
\label{L}
\end{equation}
where entropy density $s(T)$ quantifies the thermal disorder of the Dirac fluid and $n_0(T)$ refers to an effective density scale, which determines the intrinsic conductivity of the electron fluid \cite{li2022hydrodynamic, li2020hydrodynamic, majumdar2025universality}. We observe a very good agreement between the experimental data and Eqn.~\ref{L}, suggesting Dirac fluid-like electron flow in all three graphene FETs (Fig.~\ref{fig2}e) near the CNP. Notably, although $\mathcal{L}/\mathcal{L}_\mathrm{WF}$ tends to increase above that expected for a Dirac fluid (Eqn.~\ref{L}) at high values of $n$, we do not observe any upturn that may suggest the restoration of Fermi-liquid like behaviour. Nonetheless, the thermal entropy density $s=en_0\sqrt{\mathcal{L}(n=0)}$ obtained from experimental values of $n_0$ and $\mathcal{L}(n=0)$ ($\mathcal{L}$ at the CNP) for the pristine graphene at $300$~K agree with theoretical estimates \cite{muller2009graphene, yudhistira2025nonmonotonic} within a factor of $2$.

\begin{figure*}
    \centering
    \includegraphics[width=1.0\linewidth]{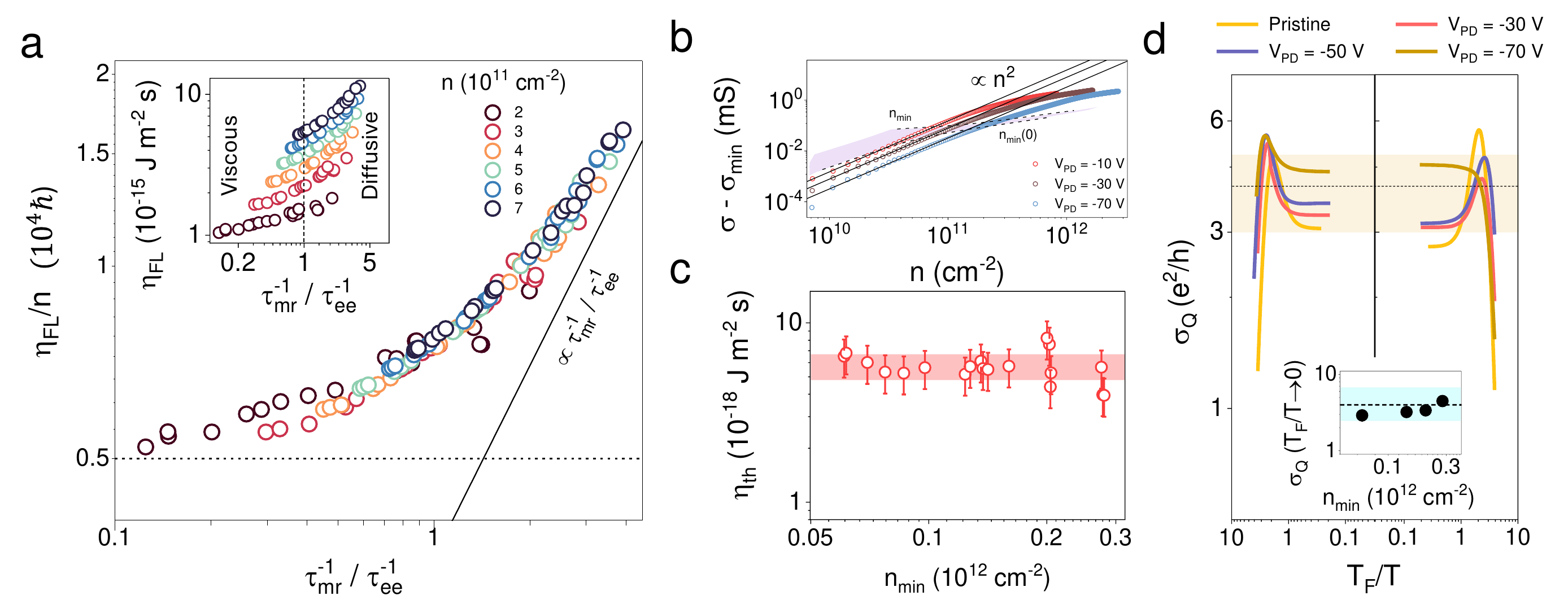}
    \caption{\textbf{Effect of disorder on shear viscosity in the Fermi liquid and thermal regimes:} (a) $\eta_\mathrm{FL}/n$ as a function of $\tau_\mathrm{mr}^{-1}/\tau_\mathrm{ee}^{-1}$ for different carrier densities, indicated using symbols of different color. The horizontal dashed line points towards the asymptotic tendency of $\eta_\mathrm{FL}/n$. In the diffusive limit, the solid line indicates $\eta_\mathrm{FL}/n \propto \tau_\mathrm{mr}^{-1}/\tau_\mathrm{ee}^{-1}$ and serves as a guide to the eye. [Inset] $\eta_\mathrm{FL}$ as a function of $\tau_\mathrm{mr}^{-1}/\tau_\mathrm{ee}^{-1}$ for different carrier densities. The vertical dotted line indicated by $\tau_\mathrm{mr}^{-1}/\tau_\mathrm{ee}^{-1} = 1$ highlights the hydrodynamic-to-diffusive crossover. (b) The plot of $\sigma - \sigma_\mathrm{min}$ as a function of $n$ for three different values of $V_\mathrm{PD}$ in device D3. The solid lines scale as $n^2$ and serve as guides to the eye. The two dashed lines denote $n_\mathrm{min}$ and $n_\mathrm{min}(0)$, with the enclosed region indicating the thermal regime being colored in blue. (c) Shear viscosity in the thermal regime ($\eta_\mathrm{th}$) as a function of $n_\mathrm{min}$ in device D3. The colored region highlights the range of variation of $\eta_\mathrm{th}$. The error bars presented here indicate the standard deviation of the variable. (d) $\sigma_\mathrm{Q}$ as a function of $T_\mathrm{F}/T$ for device D3 at different values of $V_\mathrm{PD}$. (Inset) $\sigma_\mathrm{Q}$ for $T_\mathrm{F}/T \rightarrow 0$ as a function of $n_\mathrm{min}$, with a colored region indicating the range of variation of the data points. The unit along y-axis is $e^2/h$. The dashed horizontal line corresponds to $4e^2/h$.}
    \label{fig4}
\end{figure*}

We next examine the $n$-dependence of $\mathcal{L}/\mathcal{L}_\mathrm{WF}$ (during the reverse sweep cycle of $V_\mathrm{G}$ - see Supplementary Section S8) for different values of $V_\mathrm{PD}$, i.e. at different levels of disorder. Remarkably, $\mathcal{L}/\mathcal{L}_\mathrm{WF}$ continues to follow Eqn.~\ref{L} across all values of $V_\mathrm{PD}$ (Fig.~\ref{fig3}a), although the width of the Lorentzian profile in $n$, characterised by $n_0$, increases with increasing $V_\mathrm{PD}$. Since $\mathcal{L}/\mathcal{L}_\mathrm{WF} \approx 1$ near $n \approx 0$, this broadening reflects a progressive restoration of the Wiedemann–Franz (WF) law as disorder in the graphene channel increases. Importantly, this transition from a regime that violates the WF law to one that adheres to it is reversible. By reducing the disorder through UV irradiation - applied after setting $V_\mathrm{PD} = 0$~V - we concurrently measure the electronic thermal conductivity across the channel. The results, shown in Fig.~\ref{fig3}b, reveal that the FWHM of the Lorentzian $n$-dependence of $\mathcal{L}/\mathcal{L}_\mathrm{WF}$ initially increases from $n_0$ to $n_0^*$ ($n_0^* > n_0$) as $V_\mathrm{PD}$ is swept from $0$ to $-40$~V, and subsequently returns to approximately $n_0$ as $V_\mathrm{PD}$ is brought back from $-40$ to $0$~V. This reversibility highlights the potential for dynamically tuning a hydrodynamic transport regime within a disordered graphene monolayer.

Further, this establishes $n_0$ as a crucial parameter that can be used to quantify the impact of disorder on the thermal transport in the graphene channel. With increasing $n_\mathrm{min}$, $n_0$ is found to increase monotonically, as shown in Fig.~\ref{fig3}c. The nature of this variation is theoretically expected to depend on the nature of electron-impurity scattering \cite{zarenia2019disorder}; however, a quantitative understanding of this dependence is still lacking.

The key feature of the data shown in Fig.~\ref{fig3}a is a clear increasing trend in $\mathcal{L}/\mathcal{L}_\mathrm{WF}$ with $n_\mathrm{min}$ at high electron and hole densities, as demonstrated in Fig.~\ref{fig3}d. This is consistent with the expectation that the introduction of electronic disorder boosts momentum relaxing scattering \cite{zarenia2019disorder, li2020hydrodynamic}, leading to the increase of $\mathcal{L}/\mathcal{L}_\mathrm{WF}$ towards $\approx 1$ - paving the way for a disorder-enabled hydrodynamic regime. Further, experimental data demonstrates a negligible variation in $\mathcal{L}/\mathcal{L}_\mathrm{WF}$ near the CNP at $T \simeq300$~K (Fig.~\ref{fig3}d). This can be attributed to the suppression of WF law violation near the Dirac point at room temperature by electron-phonon scattering and the presence of disorder in the system leads to a minor increase in $\mathcal{L}/\mathcal{L}_\mathrm{WF}$ at very high $n_\mathrm{min}$, which can be explained within the conventional Fermi liquid framework \cite{tu2023wiedemann, lucas2016transport}.

\section{Discussion}

Our experimental strategy to dynamically tune the momentum-relaxing scattering in the same device allows us to explore some of the key features in viscous transport as a function of disorder in a quantitative manner. An important parameter is the shear viscosity ($\eta$), which varies differently as a function of carrier density in the thermal regime ($\eta = \eta_\mathrm{th}$) and in the Fermi liquid regime ($\eta = \eta_\mathrm{FL}$).

We first focus on the Fermi liquid regime given by $n \gtrsim 10^{11}$~cm$^{-2}$ and $l_\mathrm{ee} \ll W$, where $T_\mathrm{F}>T$, and  $\eta_\mathrm{FL} = n^2e^2W^2/12\sigma$ \cite{guo2017higher,krishna2017superballistic} represents the viscous friction among the constituent particles of a mono-component charged fluid, composed of either electrons or holes. In Fig.~\ref{fig4}a (inset), $\eta_\mathrm{FL}$ has been plotted as a function of $\tau_\mathrm{mr}^{-1}/\tau_\mathrm{ee}^{-1}$ at different carrier densities. With increasing disorder, $\eta_\mathrm{FL}$ varies weakly in the viscous limit and starts increasing monotonically for $\tau^{-1}_\mathrm{mr} > \tau^{-1}_\mathrm{ee}$ across the experimental range ($1-7 \times 10^{11}$~cm$^{-2}$) of $n$. Although $\eta_\mathrm{FL}$ in a disordered system has never been precisely defined, previous works \cite{chen2022viscosity, burmistrov2019dissipative} have predicted electron-impurity scattering to enhance viscous effects. In the presence of charged impurity scattering ($\tau^{-1}_\mathrm{mr} \propto \sqrt{n}$), and noting that $\eta_\mathrm{FL} = nm\nu$ where $m = \hbar k_\mathrm{F}/v_\mathrm{F}$ and $\nu = \frac{1}{4} v^2_\mathrm{F}\tau_\mathrm{ee}$ are, respectively, the effective mass and kinematic viscosity, we find, 

\begin{equation}
    \dfrac{\eta_\mathrm{FL}}{n} = \dfrac{\hbar^3 v_\mathrm{F}^2}{n_\mathrm{d}V_\mathrm0^2}\dfrac{\tau^{-1}_\mathrm{mr}}{\tau^{-1}_\mathrm{ee}}
    \label{imp}
\end{equation}
where the electron-impurity scattering rate is given by $\tau^{-1}_\mathrm{mr} = \left(\frac{n_\mathrm{d}V_0^2 \sqrt{\pi}}{4\hbar^2 v_\mathrm{F}}\right)\sqrt{n}$ \cite{dassarma2011electronic}, $n_\mathrm{d}$ being the short-ranged neutral impurity concentration and $V_0$ being the strength of the impurity scattering potential.

For $n \gtrsim 10^{11}$~cm$^{-2}$, we observe that $\eta_\mathrm{FL}/n$ collapses onto a universal behaviour when plotted against $\tau^{-1}_\mathrm{mr}/\tau^{-1}_\mathrm{ee}$, exhibiting a linear increase at large values, in excellent agreement with Eqn.~\ref{imp} for both electrons (Fig.~\ref{fig4}a) and holes (See Supplementary Section S9). The slope of the data for $\tau^{-1}_\mathrm{mr}/\tau^{-1}_\mathrm{ee} \gg 1$ reveals the impurity scattering potential $V_0$ to be $\approx 0.03$~eV nm$^{-2}$, which is very close to theoretical estimates \cite{davoudiniya2024influence}. The physical mechanism underlying this linear increase is, however, not well understood.

On the other hand. as the system approaches the clean limit ($\tau_\mathrm{mr}^{-1}/\tau_\mathrm{ee}^{-1} \ll 1$), the shear viscosity of the graphene channel is given by $\eta_\mathrm{FL} = nm\nu =\hbar\sqrt{\pi} n^{3/2} (\nu/v_\mathrm{F})$. Now, for the pristine graphene channel, $\nu \propto 1/\sqrt{n}$ (See Supplementary Section S10), which is comparable to the decreasing trend of $\nu$ on $n$ observed in previous experimental results \cite{bandurin2016negative, talanov2024observation}. Finally, we observe in Fig.~\ref{fig4}a, that the shear viscosity per unit carrier, indicated by $\eta_\mathrm{FL}/n$, exhibits an asymptotic saturation to $\approx 5000 \hbar$, as theoretically expected from a clean Fermi liquid \cite{principi2016bulk, yudhistira2025nonmonotonic}. 

Close to the Dirac point, electrons and holes collide with each other at the Planckian rate \cite{bruin2013similarity}, which is given by $\tau^{-1}_\mathrm{P} = \alpha^2k_\mathrm{B}T/\hbar$ and is solely determined by the temperature $T$ of the electronic system. The shear viscosity of these thermally excited electrons and holes is evaluated using the expression $\eta_\mathrm{th} = e^2v_\mathrm{F}l_{mr}\tau_\mathrm{P}/2A_\mathrm{th}(T)$, where $A_\mathrm{th}(T)$ is the coefficient of $n^2$ in the $n$-dependence of $\sigma$ \cite{majumdar2025universality, lucas2016transport}.

In the thermal regime, the fluidic nature of electronic motion results in a $n^2$ dependence of the electrical conductivity $\sigma$, as shown in Fig.~\ref{fig4}b. However, the addition of disorder to ultra-clean graphene barely affects $\eta_\mathrm{th}$, as seen in Fig.~\ref{fig4}c. With increasing $n_\mathrm{min}$,  the thermal shear viscosity stays constant at $\eta_\mathrm{th} \simeq6 \times 10^{-18}$~J m$^{-2}$~s. This can be explained on the basis of the origin of this viscous motion, caused entirely by the thermally excited electrons and holes ($n_\mathrm{th}$) near the CNP. As a result, when disorder is incorporated into the system, it changes $n_\mathrm{min}(0)$ and thus $n_\mathrm{min}$, but $n_\mathrm{th}$ remains constant, thereby forcing $\eta_\mathrm{th}$ to remain constant as well. 

The inability of UV-activated disorder to affect the thermal properties of the ultra-clean graphene near the Dirac point suggests that the quantum critical behaviour of graphene in this regime remains robust against the disorder created by the radiation. This is confirmed in the plot of $\sigma_\mathrm{Q}$, evaluated using $\sigma_\mathrm{Q} = \kappa_\mathrm{e}\sigma/v_\mathrm{F}l_\mathrm{mr}s_\mathrm{th}$ \cite{majumdar2025universality, lucas2016transport} with $s_\mathrm{th}$ being the thermal entropy density \cite{yudhistira2025nonmonotonic, muller2009graphene, li2022hydrodynamic}, as a function of $T_\mathrm{F}/T$ (Fig.~\ref{fig4}d) We clearly see the saturation of $\sigma_\mathrm{Q}$, close to $T_\mathrm{F}/T \to 0$ for different disorder densities, near a universal constant with the magnitude varying around $(4 \pm 1)$~$e^2/h$ (inset of Fig.~\ref{fig4}d), in agreement with theoretical expectations \cite{muller2009graphene, majumdar2025universality}.

In conclusion, our experimental results have successfully established a non-invasive tool, comprising a combination of UV radiation and gate electric field, which is capable of reversibly tuning the disorder level in a high-quality hBN-encapsulated graphene monolayer with high precision. This control over the introduction of disorder in a clean graphene FET facilitates the demonstration of a dynamically tunable charged fluid, which was confirmed by thermal noise spectroscopy of the disordered Dirac fermions navigating through the UV-irradiated graphene layer. Additionally, our results assess the impact of charged disorder on electronic viscosity in monolayer graphene, offering new insights into both the Dirac and Fermi liquid phases. Our findings act as the testing bed for several existing theoretical models of relativistic hydrodynamics, focusing on the interplay of momentum-conserving and momentum-relaxing scattering to explain the onset of electron fluidity in monolayer graphene. 

\section{Acknowledgment}
The authors gratefully acknowledge the usage of the MNCF and NNFC facilities at CeNSE, IISc. The authors would also like to acknowledge fruitful discussions with N. Chaddha and S. Mukerjee. A.G. acknowledges financial support from J. C. Bose Fellowship and a project under NanoMission, Department of Science and Technology, India. Ak.G. thanks the Ministry of Education, Govt. of India, for the Prime Minister's Research Fellowship (PMRF). K.W. and T.T. acknowledge support from the JSPS KAKENHI (Grant Numbers 21H05233 and 23H02052), the CREST (JPMJCR24A5), JST and World Premier International Research Center Initiative (WPI), MEXT, Japan.

\bibliography{ref}

\end{document}


\title{Dynamically tunable hydrodynamic transport in boron nitride-encapsulated graphene: (Supplementary Information)}

\author{Akash Gugnani} \thanks{equal contribution}
\email{akashgugnani@iisc.ac.in}
\affiliation{Department of Physics, Indian Institute of Science, Bangalore 560012, India}
\author{Aniket Majumdar} \thanks{equal contribution}
\email{aniketm@iisc.ac.in}
\affiliation{Department of Physics, Indian Institute of Science, Bangalore 560012, India}
\author{Kenji Watanabe}
\affiliation{Research Center for Electronic and Optical Materials, National Institute for Materials Science, 1-1 Namiki, Tsukuba 305-0044, Japan}
\author{Takashi Taniguchi}
\affiliation{Research Center for Materials Nanoarchitectonics, National Institute for Materials Science, 1-1 Namiki, Tsukuba 305-0044, Japan}
\author{Arindam Ghosh}
\email{arindam@iisc.ac.in}
\affiliation{Department of Physics, Indian Institute of Science, Bangalore 560012, India}
\affiliation{Center for Nano Science and Engineering, Indian Institute of Science, Bangalore 560012, India}
\maketitle  

\section{Device fabrication}

The hBN-encapsulated monolayer graphene heterostructures were fabricated using a dry transfer technique. All constituent flakes in the van der Waals (vdW) stack were mechanically exfoliated and assembled using a hemispherical polydimethylsiloxane (PDMS, 10:1 base-to-curing-agent ratio) stamp coated with a thin film of propylene carbonate (PC) \cite{xie2017facile}. The assembled heterostructure was transferred onto a pre-patterned Si$^{++}$/SiO$_2$ substrate at $180^\circ$~C. Electrical contacts to the graphene layer were defined via electron beam lithography (EBL), followed by reactive ion etching (RIE) and subsequent thermal evaporation of Cr/Au ($5/50$~nm) \cite{wang2013one}. Cleanliness of the device was further ensured by Raman peaks \cite{ferrari2006raman, lin2011clean} of the Gr flake and thermal annealing \cite{engels2014impact} of the stack in an inert (Ar/H$_278$) environment at $200^\circ$~C inside a glove box.

\section{Device cleanliness}

\begin{figure}[tbh!]
    \centering
    \hspace*{-0.2cm}
    \vspace{0cm}
    \includegraphics[scale=0.5]{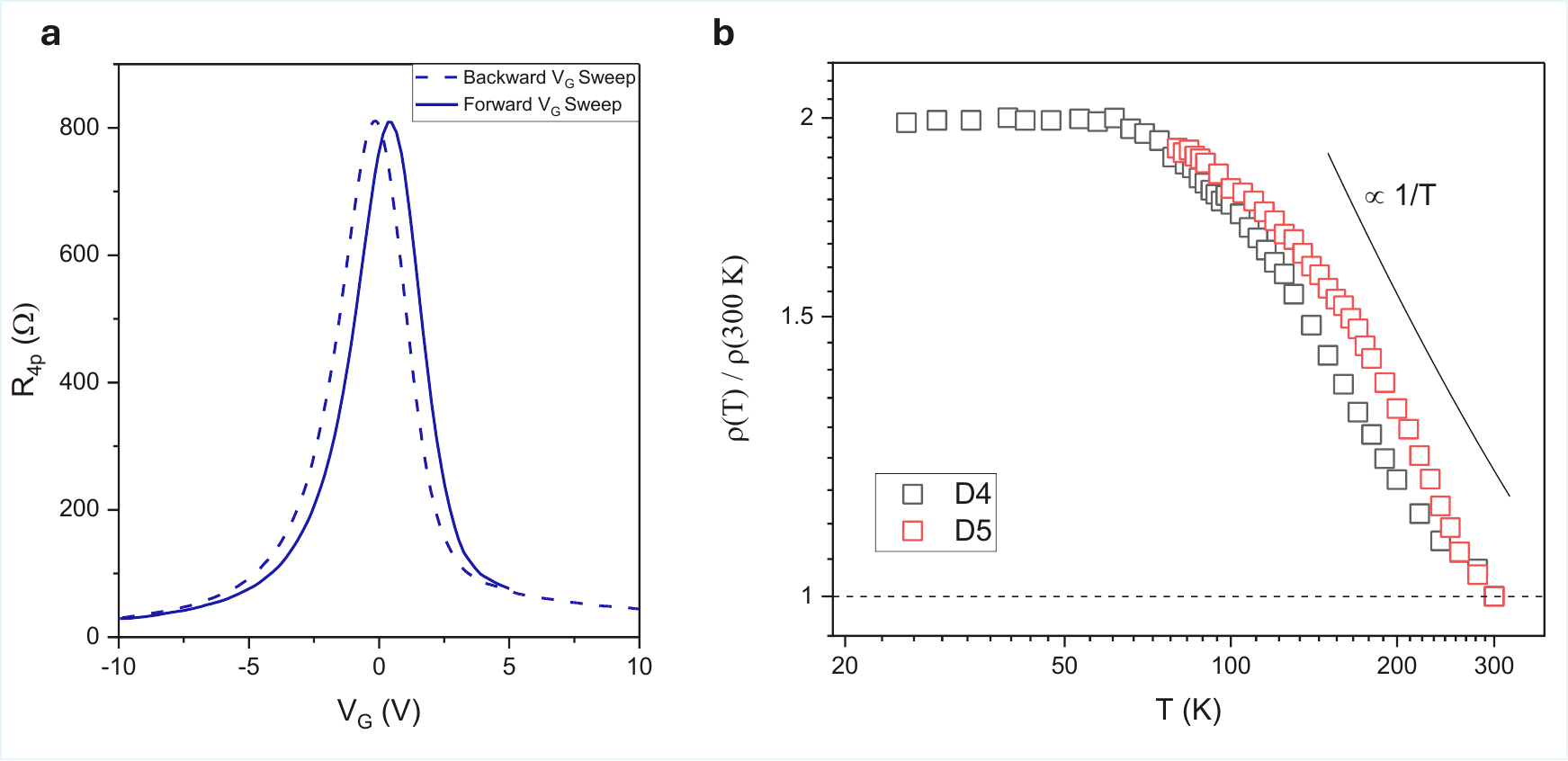}
    \caption{\textbf{Extraction of induced bandgap in hBN-encapsulated graphene:} (a) Electrical transfer characteristics of device D5 at room temperature.  (b) Variation of normalized resistivity $\rho(T)/\rho(300K)$ with temperature $T$ at the charge neutrality point for devices D4 and D5, plotted on a logarithmic scale. Black solid line is $\propto 1/T$ and serves as a guide to the eye. Red and black hollow squares show the experimental data for devices D4 and D5 respectively. }
    \label{R1Q1}
\end{figure}

The transfer characteristics ($R_\mathrm{4p}-V_\mathrm{G}$) of the device were first recorded prior to UV exposure, as shown for device D5 in Fig.~\ref{R1Q1}a. The charge neutrality point (CNP) was located at ($V_\mathrm{D}=-1.0$~V), and negligible hysteresis ($\Delta V=0.7$~V) was observed between forward and backwards $V_\mathrm{G}$ sweeps. This indicates minimal extrinsic doping and suggests that the device is mostly free from intrinsic or interfacial trap states, which typically contribute to hysteresis. 

Boron nitride-encapsulated high mobility graphene samples are also susceptible to the formation of a small band gap at the charge neutrality point, which might originate due to unintentional graphene-hBN alignment-induced moir\'e superlattice or even strain-induced structural relaxation of the graphene lattice \cite{jung2015origin}. However, during the fabrication of our samples, we have taken additional care to prevent any graphene-hBN alignment - as a result, the transfer characteristics do not show any features of moir\'e physics at room temperature. Further, we performed low-temperature transport measurements in our devices to confirm the gapless nature of the band structure near the charge neutrality point (CNP). As shown in Fig~\ref{R1Q1}b for two of our devices, $\rho$ decreases roughly by a factor of two as $T$ is increased from $20$~K to $300$~K. This $1/T$-dependent variation of resistivity $\rho$ can be attributed to the gapless nature of the Dirac cone or the viscous electron fluid in graphene near the Dirac point, indicating the absence of any small band gap. Even if we consider that this monotonically decreasing variation of $\rho$ with increasing $T$ is entirely due to the opening of a small band gap at the CNP, we can fit the temperature-dependent resistivity $\rho(T)$ near the CNP with the thermally-activated transport equation given by 

\begin{equation} \label{bandgap}
    \rho(T)/\rho_\mathrm{0} = \exp{(\Delta/k_\mathrm{B}T)}
\end{equation}

The extracted band-gaps $\Delta = 9-14$~meV are still small compared to the thermal energy of electrons at room temperature ($k_\mathrm{B}T \simeq 25$~meV at $300$~K), confirming that any strain-induced gap in the density of states is too small to impact our experiments, which were carried out at room temperature.

\section{Experimental circuit for performing UV-mediated disorder-doping}

\begin{figure}
    \centering
    \hspace*{0cm}
    \includegraphics[scale=0.7]{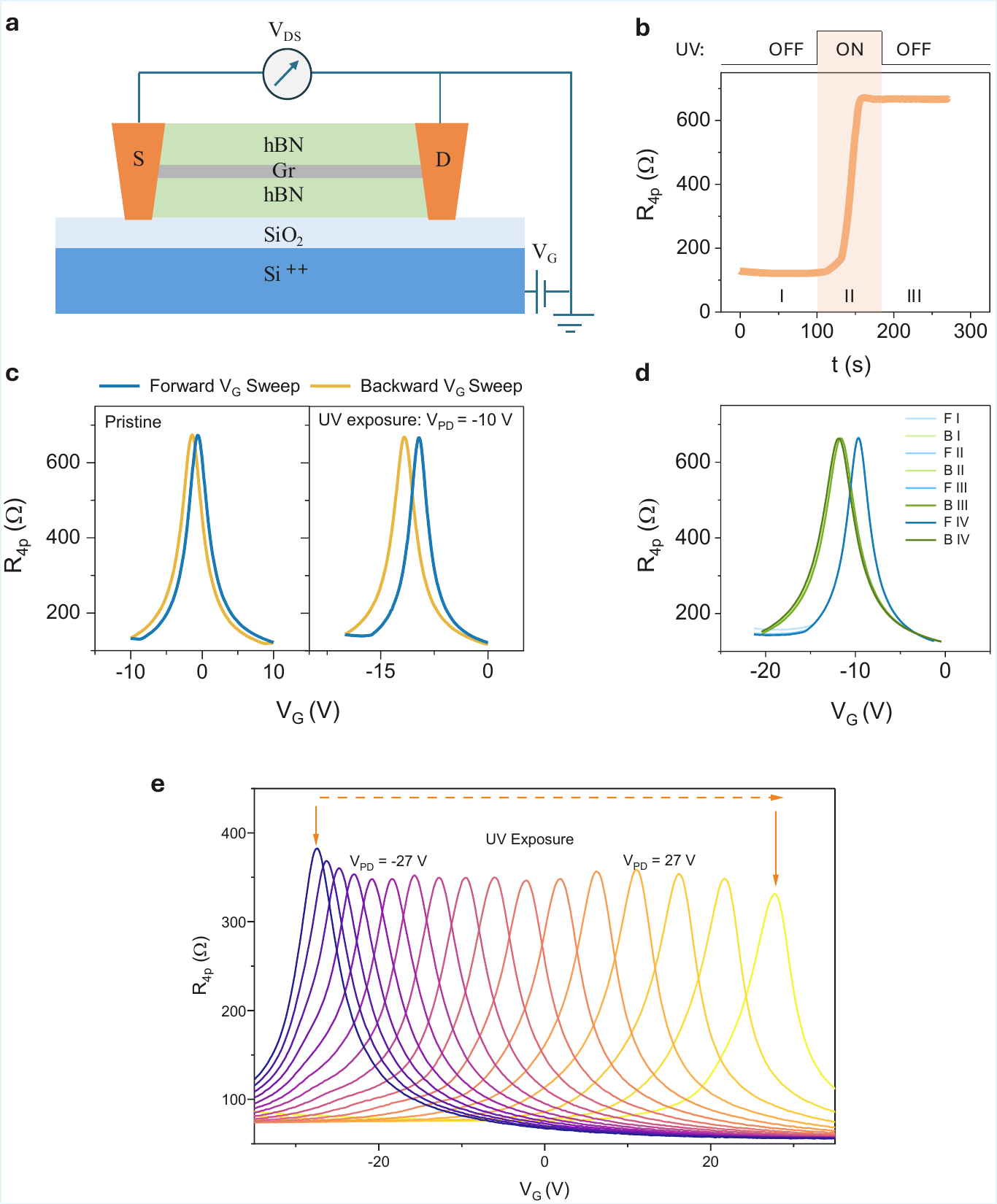}
    \caption{\textbf{Electrical transport measurement scheme of hBN encapsulated graphene device subjected to UV irradiation} (a) Schematic of the hBN/Gr/hBN heterostructure on a Si$^{++}$/SiO$_2$ substrate. S and D represent the source and drain gold electrodes, which are edge-contacted to the graphene channel and used to measure $V_\mathrm{DS}$ using a lock-in amplifier in a four-terminal contact configuration (not shown in the figure). (b) Variation of $R_\mathrm{4p}$ with ${t}$ shows the channel resistance before (region I), during (region II), and after (region III) the UV exposure at $V_\mathrm{PD}=-10$~V. The shaded part represents UV LED ON and OFF status.  (c) $R_\mathrm{4p}-V_\mathrm{G}$ of the device before (left panel) and after (right panel) UV exposure at $V_\mathrm{{PD}}=-10$~V, causing a shift in CNP and an increase in hysteresis. (d) Four cycles of $R_\mathrm{4p}-V_\mathrm{G}$ of the device taken after UV exposure at $V_\mathrm{PD}=-10$~V. F and B in the figure refer to forward and backward $V_\mathrm{G}$ sweep directions, respectively. Also, the forward and backwards $V_\mathrm{G}$ sweeps individually overlap over each other across the multiple cycles and hence are difficult to isolate. (e) Electrical transport characteristics of device D4 at 300 K. $R_\mathrm{4p}-V_\mathrm{G}$ curves show UV-induced doping, with the charge neutrality point shifting from -27 V (initial, blue) to +27 V (final, yellow). Intermediate measurements (colour gradient) track the progressive CNP shift marked by the direction of the arrow.} 
    \label{fig:device}
\end{figure}

Electrical transport measurements were performed at room temperature ($\simeq 296$~K) under high-vacuum conditions ($\sim  10^{-6}$~mbar). A schematic of the measurement setup is shown in Fig.~\ref{fig:device}a. The device consists of monolayer graphene encapsulated between hBN layers, fabricated on a p-doped Si$^{++}$/SiO$_2$ substrate, where the $285$~nm thick SiO$_2$ layer serves as the back-gate dielectric. The back-gate voltage ($V_\mathrm{G}$) was applied through the doped silicon substrate using a Keithley 2400 SourceMeter. Electrical contacts were patterned in a four-probe configuration, with source (S) and drain (D) electrodes forming one-dimensional edge contacts to the graphene channel. The four-probe resistance ($R_\mathrm{4p}$) was measured using a SR830 lock-in amplifier. For photodoping, a $235$~nm UV LED (1125-MTE2350D-UV-ND), was used to execute blanket illumination of the device. The LED was driven at a constant voltage bias of $V_\mathrm{LED}\simeq5$~V using a DC power supply, corresponding to an optical power density of $\approx 1$~pW/~$\mu$m$^2$. The UV LED illumination was used with a convex lens to ensure spatially uniform exposure and performed at various $V_\mathrm{G}$ for controlled photodoping experiments. 

\par \indent Photodoping of the graphene channel was performed by illuminating the device with the $235$~nm UV LED, biased at a constant voltage $V_\mathrm{LED}$, while simultaneously applying a fixed back-gate voltage $V_\mathrm{PD}$, referred to as the photodoping voltage. $R_\mathrm{4p}$ of the channel was continuously monitored in real time ($t$) during the UV exposure ($\simeq 100$~s), as shown in Fig.~\ref{fig:device}b. The three regions, labelled I, II, and III in Fig.~\ref{fig:device}b, correspond to the pre-illumination, illumination, and post-illumination stages, respectively. As shown in region II of Fig.~\ref{fig:device}b, when the device was subjected to irradiation, UV-induced charge carriers originating from defect states in the hBN flake and the hBN/SiO$_2$ interface migrate into or out of the graphene channel under the influence of the applied gate electric field. The remaining immobile residual ions within the hBN flake and at the hBN/SiO$_2$ interface form a screening layer that suppresses further charge transfer to the graphene. As a result, after a certain time, the carrier density in the graphene channel stabilises, and the four-probe resistance $R_\mathrm{4p}$ saturates to a constant value. In this case, with  $V_\mathrm{PD}=-10$~V, photogenerated electrons are driven toward the graphene channel, while the residual immobile positive ions form a screening layer within the hBN and at the hBN/SiO$_2$ interface. This screening effectively neutralises the applied negative gate voltage bias. The resistance saturation timescale in this case was $~\sim100$~s. Once $R_\mathrm{4p}$ saturated, both $V_\mathrm{G}$ and $V_\mathrm{{LED}}$ were gradually tuned to $0$~V. The net UV-induced photodoping caused a pronounced shift of the CNP from ${V_\mathrm{D}} \simeq 0$~V (Fig.~\ref{fig:device}c: left panel) to ${V_\mathrm{{D}}} \simeq -10$ V (Fig.~\ref{fig:device}c: right panel). To evaluate the temporal stability of this doping state, transfer characteristics of the UV-exposed graphene FET were continuously monitored over several days. As shown in Fig.~\ref{fig:device}d, repeated forward and backwards $V_\mathrm{G}$ sweeps revealed negligible variation in the position of the CNP over a period of $4-5$ days. These observations demonstrate the persistence of the photodoping mechanism under the experimental conditions. Additionally, we performed $15$ consecutive photodoping cycles, shifting the CNP from $0$~V to $-30$~V in steps of $2$~V per cycle. Across these repeated ON/OFF UV cycles, we found no evidence of irreversible degradation or failure to reset the CNP, as shown in Fig.~\ref{fig:device}e, for one such device. These results strongly support the long-term stability and robust nature of the graphene-hBN interface when subjected to UV exposure, at least within the experimental timescales and illumination conditions used in our study.

\begin{figure}[tbh!]
    \centering
    \hspace*{-0.2cm}
    \vspace{0cm}
    \includegraphics[scale=1]{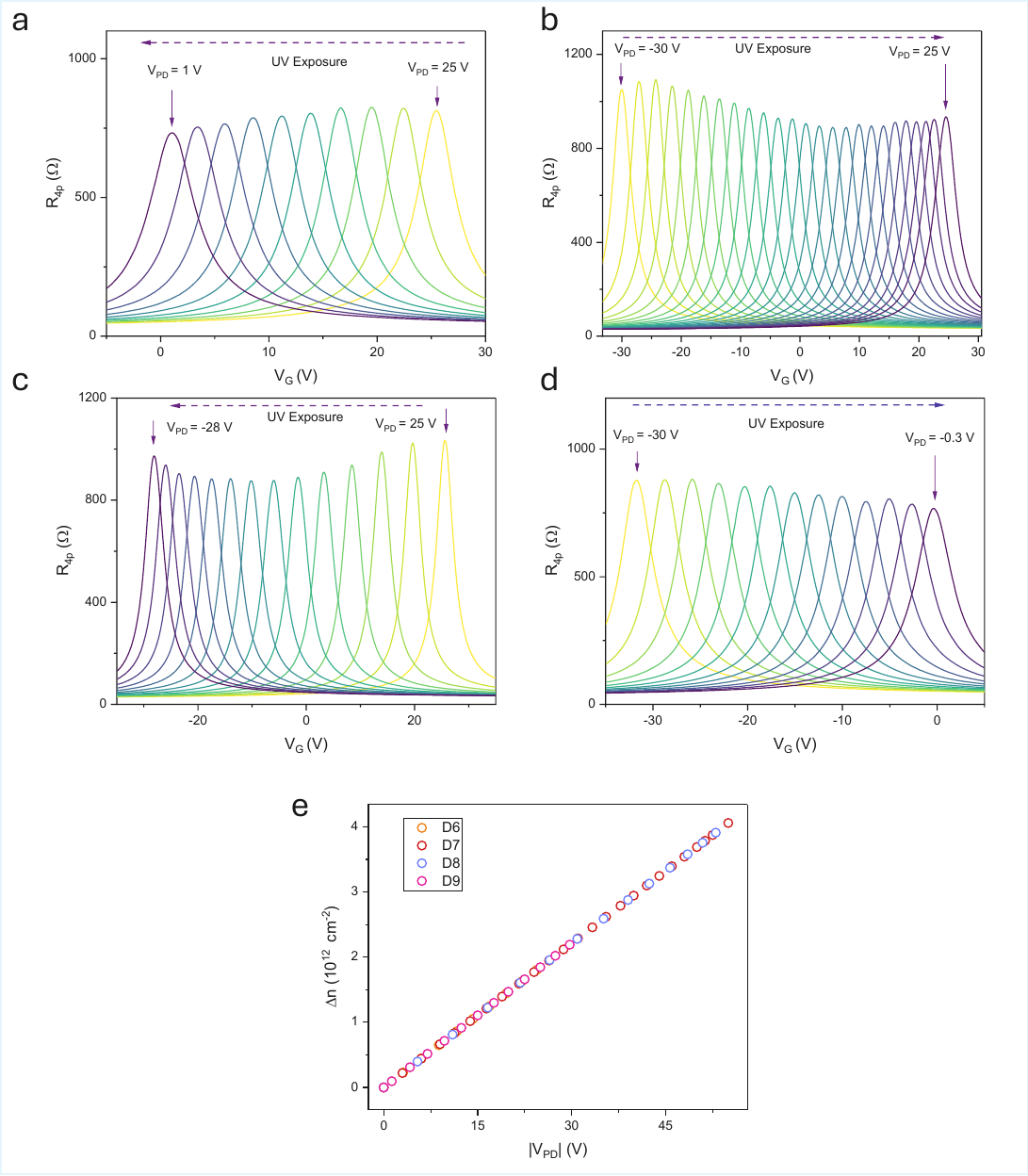}
    \caption{\textbf{Controlled and reproducible shift of charge neutrality point across devices:} (a-d)  Resistance versus gate voltage ($R_\mathrm{4p}-V_\mathrm{G}$) for four additional hBN-encapsulated graphene devices, showing tunable CNP positions achieved via controlled UV doping. The dotted arrow shows the direction of the CNP shift, with the initial and final positions marked on top. (e) Variation of $\Delta n$ (density of charge carriers introduced by photo-doping) as a function of $V_\mathrm{PD}$. The colour coding of the hollow circle corresponds to different devices as shown in (a-d). }
    \label{R1Q3}
\end{figure}

Further, to establish the robustness of our UV-photodoping technique, we systematically characterised four additional hBN-encapsulated graphene devices (D6-D9) (Fig.~\ref{R1Q3}a-d). All devices exhibited linear shifts of the CNP proportional to the applied $V_\mathrm{PD}$ (Fig.~\ref{R1Q3}e). The transfer curves post-UV exposure(s) confirm uniform electron doping, with device-to-device variations.

\subsection{Advantages of using UV radiation over visible light for defect generation}

\begin{figure}[tbh!]
    \centering
    \hspace*{-0.2cm}
    \vspace{0cm}
    \includegraphics[scale=0.54]{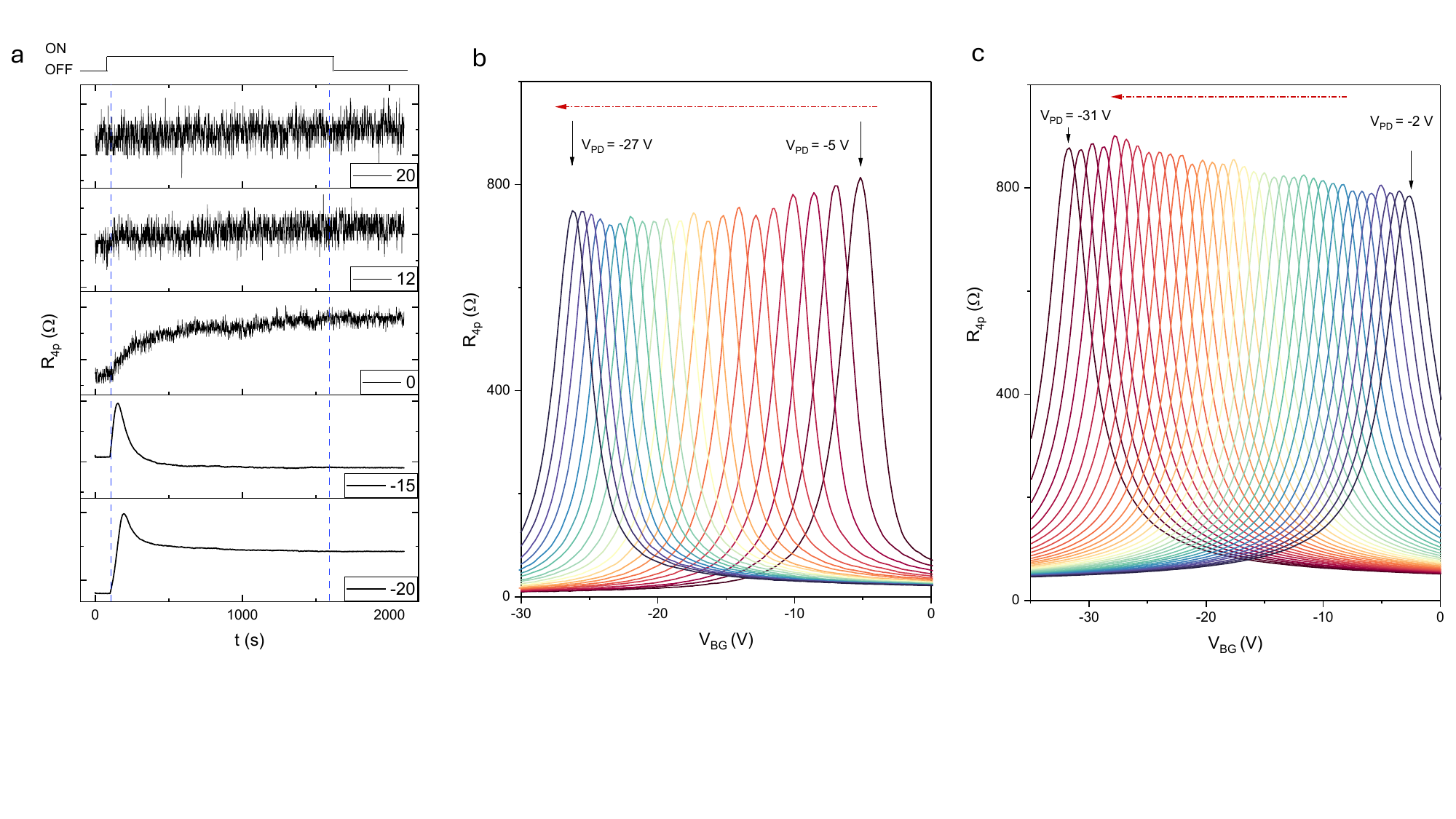}
    \caption{\textbf{Defect generation using visible light:} (a) Time evolution of channel resistance $R_\mathrm{4p}$ under UV exposure at different photodoping voltages ($V_\mathrm{PD}$ = $20$~V, $12$~V, $0$~V, $-15$~V and $-30$~V, top panel to bottom panel). Blue dotted lines indicate UV LED ON/OFF time with the symbol presented on top. (b-c) Resistance versus gate voltage ($R_\mathrm{4p}-V_\mathrm{G}$) characteristics for devices D4 (b) and D5 (c). Arrows indicate the direction of charge neutrality point shift under (b) 405 nm and (c) 532 nm illumination. Different colors represent intermediate $R_\mathrm{4p}-V_\mathrm{G}$ sweeps taken during visible light doping at intervals of ten seconds. }
    \label{R2Q1}
\end{figure}

Similar experiments, reported in previous studies, have demonstrated photo-doping using visible light (with photon energies varying between $1.5 - 2.0$~eV). This can induce photo-doping in graphene/hBN heterostructures by exciting shallow in-gap defect states in hBN, such as carbon substitutional defects and boron vacancies \cite{attaccalite2011coupling, zhang2017two}, which typically lie $\sim 1.5 - 2.5$~eV above the valence band maximum. However, excitation of these shallow states is not sufficient to generate the level of charge disorder necessary for investigating the full effect of disorder in viscous electron flow in graphene. For disorder doping beyond $10^{12}$~cm$^{-2}$, access to deep defect states (near the conduction band minimum), oxygen/nitrogen vacancies, and trap states at the hBN/SiO$_\mathrm{2}$ interface is required. All these states lie at energies $> 3.5-4.5$~eV above the hBN valence band maximum, and this can be facilitated only using higher-energy UV photons. We have tried using lower photon energies ($405$~nm and $532$~nm) as well; however we observed that these photons have two shortcomings with regards to this specific experiment - (a) shifting the CNP requires a long time (Fig.~\ref{R2Q1}a), even under the application of significantly high photon intensities ( $> 500$~fW/$\mu$m$^2$), and (b) negligible disorder was introduced on shifting the CNP of graphene to higher (positive or negative) gate voltages i.e. for increasing values of |$V_\mathrm{PD}$|, as shown in Figs.~\ref{R2Q1}b and c.

\section{Signatures of UV-induced photodoping on charge transport in boron nitride-encapsulated graphene devices}

\subsection{Electrical transfer characteristics of UV-doped graphene}

UV exposure, accompanied by the application of a negative $V_\mathrm{G}$ bias, facilitates photo-induced electron doping in graphene, which is attributed to the donor-like defects \cite{attaccalite2011coupling, zhang2017two, ju2014photoinduced} within the hBN flake, and across the vdW interfaces. These donor states release electrons upon UV illumination, which are driven towards the graphene layer by the applied back-gate electric field. The remnant immobile positive ions reside within the hBN layer and at the hBN/SiO$_2$ interface, forming a layer of localised charges that screen $V_\mathrm{G}$ and pin the CNP close to the applied $V_\mathrm{PD}$ \cite{neumann2016spatial}. Fig.~\ref{fig:sweep_mobility}a depicts the transfer characteristics during the forward and backwards $V_\mathrm{G}$ sweeps, obtained by UV exposure for different values of $V_\mathrm{PD}$, in the top panel and bottom panel, respectively. The back-gate voltage effectively determines the photodoping voltage $V_\mathrm{PD}$, responsible for charge transfer. The UV-induced electron doping in the graphene channel can be finely tuned by varying $V_\mathrm{PD}$. 

\subsection{Precise control of disorder via UV-photodoping}

Our technique has demonstrated a highly precise disorder-doping modulation (Fig.~\ref{fig:sweep_mobility}b), with a resolution of $3.3\times10^9$~cm$^{-2}$~/V indicating that the Fermi energy can be shifted to as low as $6.5$~meV for changing $V_\mathrm{PD}$ by $1$~V. Further, our experimental observations have recorded an addition of charged inhomogeneity $\approx 3\times10^{11}$~cm$^{-2}$ without causing irreversible damage to the graphene FET. This enables precise positioning of the CNP, allowing it to be pinned at virtually any desired gate voltage within the experimental range.

\subsection{Carrier mobility degradation under photo-induced doping}

\begin{figure}
    \centering
    \hspace*{-0.5cm}
    \includegraphics[scale=0.75]{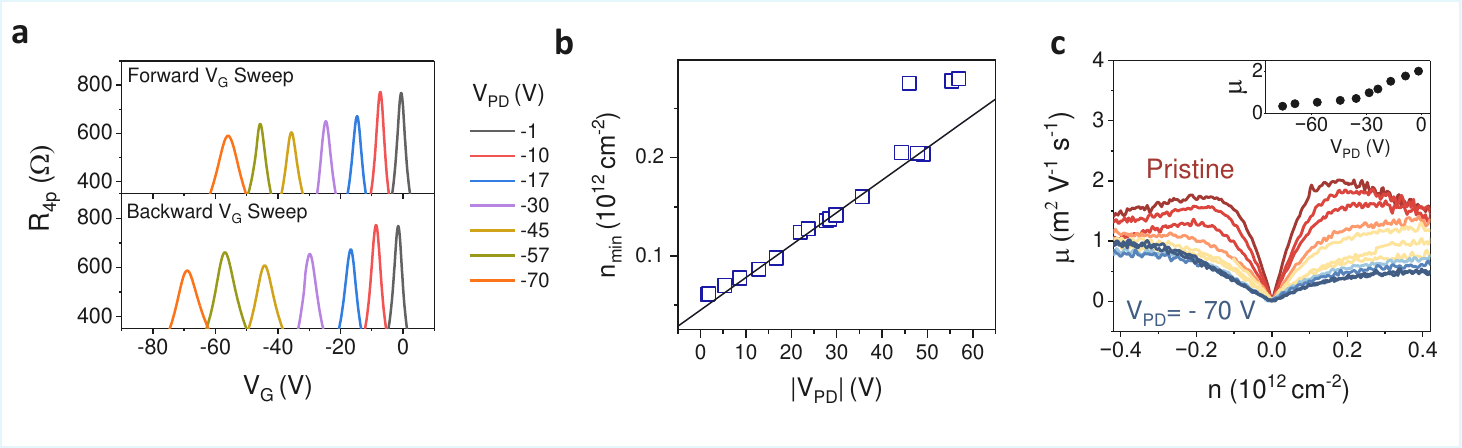}
    \caption{\textbf{UV-induced electrical transport characteristics in hBN-encapsulated graphene device} (a) Electrical transfer characteristics of device D3, for different degrees of photo-induced electron doping, as a function of $V_\mathrm{G}$, for both forward and backwards gate voltage sweeps. The different colours correspond to different values of $V_\mathrm{{PD}}$ as shown in the table to the right, and the same colour scheme has been used for both top and bottom panels. For clarity, the black curve corresponds to pristine graphene, and the orange curve corresponds to $V_\mathrm{{PD}}=-70$~V. (b) Variation of $n_\mathrm{min}$ with $|V_\mathrm{PD}|$. Blue squares are data points taken at individual $V_\mathrm{PD}$ with UV exposure, and the solid black line serves as a guide to the eye and scales linearly with $|{V_\mathrm{PD}}|$. (c) Carrier mobility ($\mu$) as a function of carrier concentration ($n$) for increasing disorder level $V_\mathrm{PD}$. Inset shows the suppression of electron mobility $\mathrm{\mu}$~(m$^2$V$^{-1}$s$^{-1}$)  with $V_\mathrm{{PD}}$ at $n=10^{11}$~cm$^{-2}$.}
    \label{fig:sweep_mobility}
\end{figure}

Carrier mobilities were extracted from the transfer curves obtained during the backwards $V_\mathrm{G}$ sweep (Fig.~\ref{fig:sweep_mobility}a: bottom panel), using the Drude formula. A systematic suppression of mobility is observed with increasing photodoping voltage $V_\mathrm{PD}$, as shown in Fig.~\ref{fig:sweep_mobility}c, indicating a progressive increase in disorder, likely caused by a higher density of UV-activated trap states. This trend is summarized in the inset of the Fig.~\ref{fig:sweep_mobility}c, where the electron mobility $\mu$ decreases by nearly an order of magnitude as $V_\mathrm{PD}$ is varied from $0$~V to $-70$~V at a carrier density of $n=10^{11}$~cm$^{-2}$. A similar suppression is observed for hole mobility as well (Fig.~\ref{fig:sweep_mobility}c),  confirming that the degradation in transport is roughly symmetric for both carrier types and is directly correlated to the UV-induced disorder concentration.

\subsection{Effect of UV photo-doping on the scattering rate}

\begin{figure}[tbh!]
    \centering
    \hspace*{-0.5cm}
    \vspace{0cm}
    \includegraphics[scale=0.55]{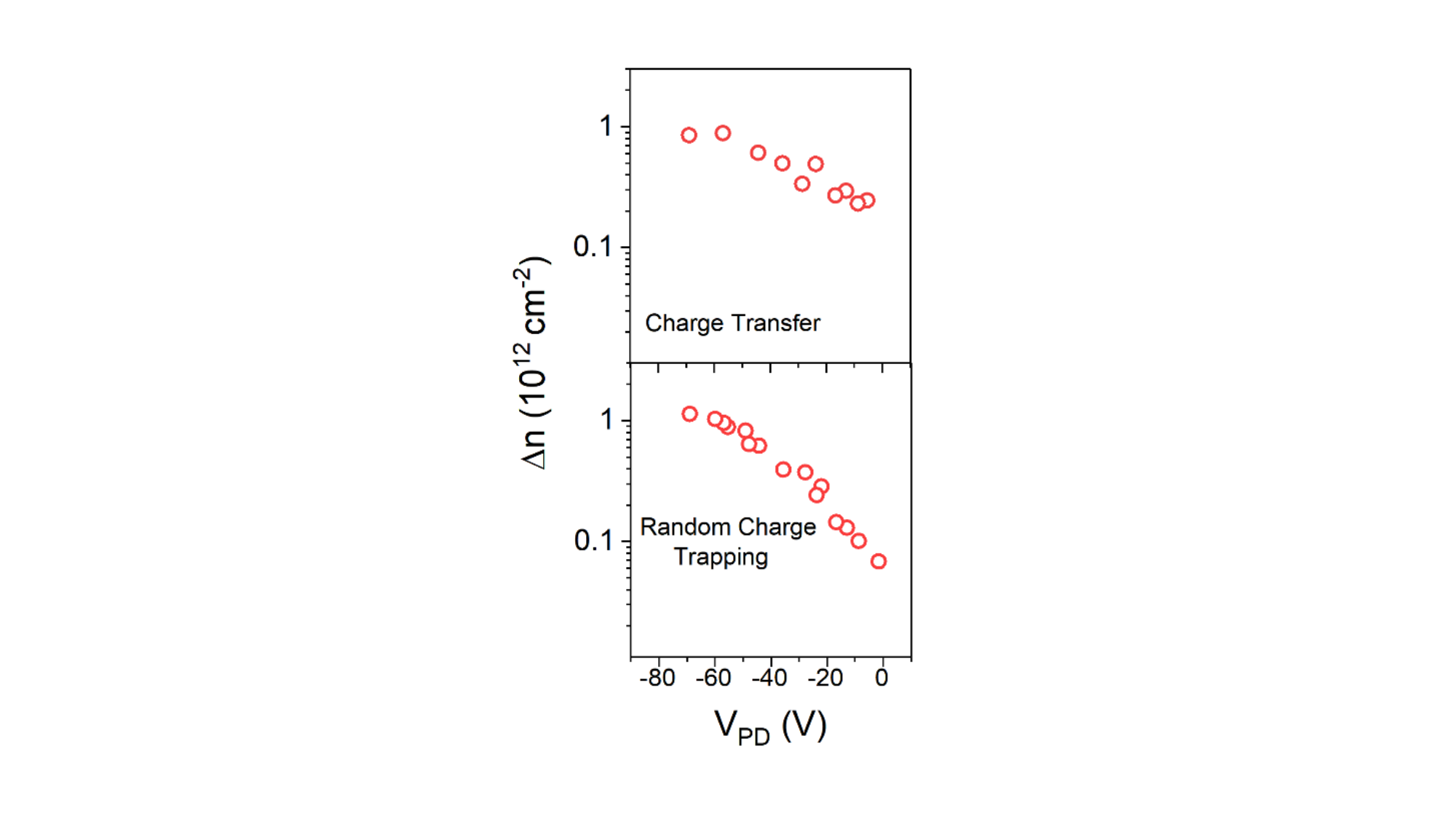}
    \caption{\textbf{Scattering mechanism involved in UV photo-doping:} Variation of $\Delta n$ as a function of $V_\mathrm{PD}$. The top panel shows charge transfer contribution (photogating) to the total carrier density change $\Delta n$, and the bottom panel accounts for random charge trapping contribution, showing disorder-induced doping variation. }
    \label{R2Q4}
\end{figure}

The interlayer charge transfer (photogating) mechanism primarily shifts the charge neutrality point by controlling the density of charge carriers within the graphene channel, without affecting the quasiparticle scattering rate at a constant carrier density. In contrast, random trapping at hBN/SiO$_2$ interfaces creates long-range scatterers ($\tau \propto \sqrt{n}$) that degrade carrier mobility. Fig.~\ref{R2Q4} quantifies these competing effects: the CNP shift can be used to calculate the charge density transferred to graphene, while an increase in hysteresis can be considered as a measure of mobile charges trapped in interfacial defect states.

\section{Reducing the UV-activated disorder concentration in the graphene FET}
We observe that prolonged UV exposure of the disorder-doped graphene channel at $V_\mathrm{PD}=0$~V effectively restores its electrical transport characteristics to its pristine condition. This recovery is marked by a shift in the CNP, carrier mobility ($\mu$) and electrical hysteresis ($\Delta V$) in $V_\mathrm{G}$ - all back restored to their initial values. In contrast, UV illumination at a non-zero photodoping voltage ($V_\mathrm{PD}\neq 0$~V) leads to reduced mobility, increased hysteretic behaviour, etc., which can be attributed to the formation of a layer of ionized impurity states in the hBN layers and at the vdW interfaces, as well as charge transfer to or from the graphene channel. The reset behaviour, observed for $V_\mathrm{PD}=0$~V in the presence of UV irradiation, is likely driven by the reversal in direction of the external gate electric field, which facilitates charge recombination and enables the de-trapping of the previously electron-doped impurity states via hole doping. This process neutralises the extrinsic residual doping, as evident by the recovery of higher mobility and minimal hysteresis, which are the hallmarks of a pristine undoped device \cite{le2024spatial}. 

\begin{figure}[tbh!]
    \centering
    \hspace*{-0.2cm}
    \vspace{0cm}
    \includegraphics[scale=0.58]{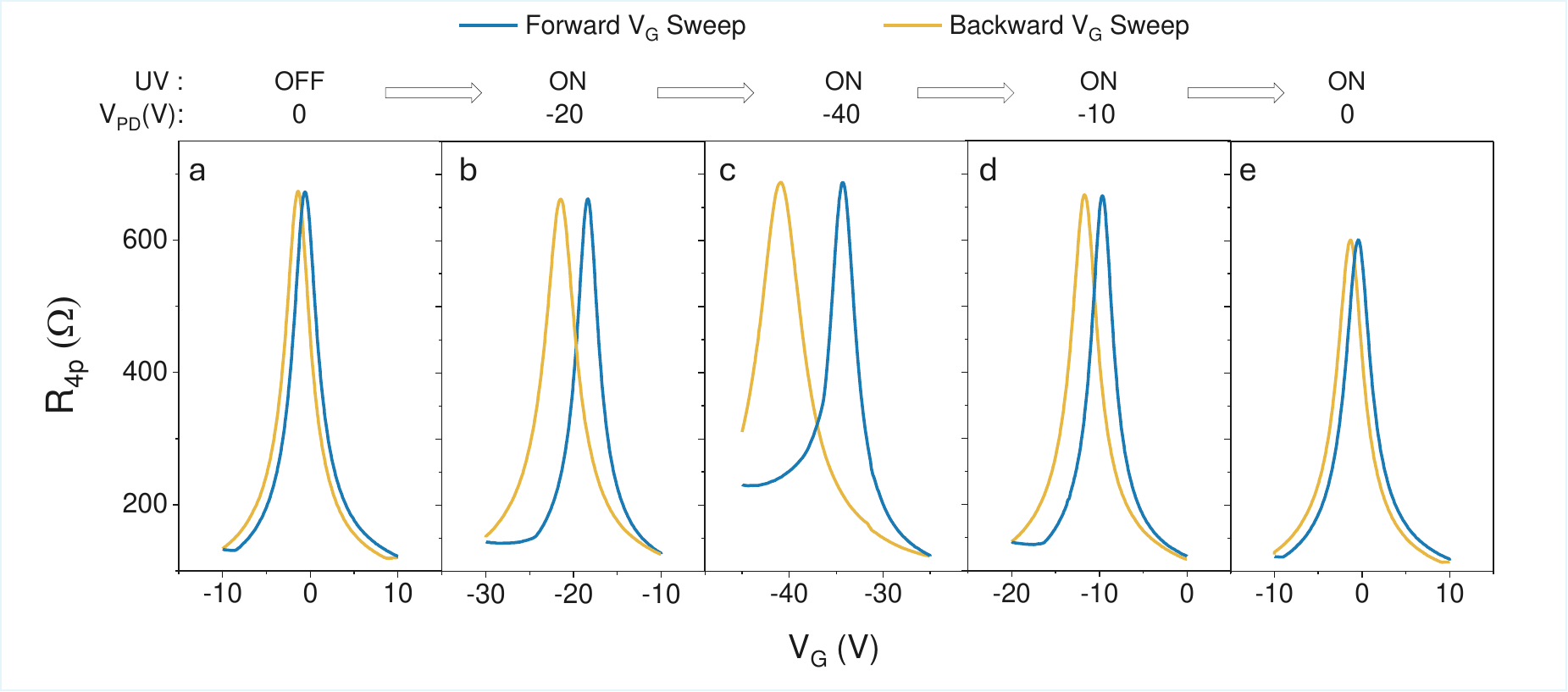}
    \caption{\textbf{UV-modulated reset of the electrical transport characteristics of the graphene channel} (a-e) $R_\mathrm{4p}-V_\mathrm{G}$ plots for the device D2 under UV exposure and varying $V_\mathrm{PD}$ bias. Fig.~\ref{shift_RVg}a-c show progressive broadening of $R_\mathrm{4p}-V_\mathrm{G}$ and increased hysteresis ($\Delta V = 0.7$~V$\to$ $3.0$~V$\to$ $7.0$~V in fig. a, b and c respectively) between forward and backward $V_\mathrm{G}$ sweeps while Fig.~\ref{shift_RVg}d-e show restoration of the device's electrical characteristics, evidenced by sharpened $R_\mathrm{4p}-V_\mathrm{G}$ transitions and reduced hysteresis ($\Delta V = 2.0$~V $\to$ $0.7$~V) in Fig.~\ref{shift_RVg}:d-e respectively. }
    \label{shift_RVg}
\end{figure}

As shown in Fig.~\ref{shift_RVg}a, the $R_\mathrm{4p}-V_\mathrm{G}$ characteristics of the pristine device ($V_\mathrm{PD}=0$~V, UV OFF) display a sharp resistance peak at the charge neutrality point ($V_\mathrm{D}=-1.0$~V) with minimal hysteresis ($\Delta V=0.7$~V) between the forward and backward  $V_\mathrm{G}$ sweeps. Upon UV exposure at increasingly negative photodoping voltages ($V_\mathrm{PD}=-20$~V and $V_\mathrm{PD}=-40$~V), shown in Figs.~\ref{shift_RVg}b–c, the CNP shifts to more negative values, indicating gradual electron doping of the graphene channel. Simultaneously, the transfer curves broaden and the hysteresis increases, which can be argued to have risen out of UV-induced trap state activation. Figs.~\ref{shift_RVg}d–e illustrate the recovery behaviour of the device following subsequent UV exposures at less negative and zero back-gate voltage biases ($V_\mathrm{PD}=-10$~V and $V_\mathrm{PD}=0$~V, respectively). These exposures lead to UV-induced extrinsic hole doping, reversing the effect of the previously induced electron doping and restoring the device to its pristine electrical state. This can be understood from the increase in sharpness of the resistance peaks and the suppression of electrical hysteresis in the transfer characteristics obtained at the end of this UV-induced hole doping. Notably, Fig.~\ref{shift_RVg}e shows that UV illumination at $V_\mathrm{PD}=0$~V fully resets the transfer characteristics to match those of the original pristine device shown in Fig.~\ref{shift_RVg}a, confirming the reversibility of the photodoping process. 

We also note that the CNP resistance under UV photodoping exhibits a non-monotonic variation, and this behavior is non-universal across devices. This can be attributed to a competition between two opposing effects: (i) increased residual carrier density ($n_\mathrm{min}$) that lowers CNP resistance on photo-doping, and (ii) mobility suppression ($\mu$) that elevates the CNP resistance through enhanced disorder scattering. For example, in device D3 (Fig.~\ref{fig:sweep_mobility}a), the 6-fold increase in $n_\mathrm{min}$ outweighs the 4-fold mobility reduction (Fig.~\ref{fig:sweep_mobility}c), yielding a net decrease in CNP resistance on increase in photo-doping. However, device D2 (Fig~\ref{shift_RVg}) exhibits the reverse trend due to its lower initial mobility ($12,000$~cm$^2$/V·s vs. $20,000$~cm$^2$/V·s).

\begin{figure}[tbh!]
    \centering
    \hspace*{-0.4cm}
    \includegraphics[scale=0.62]{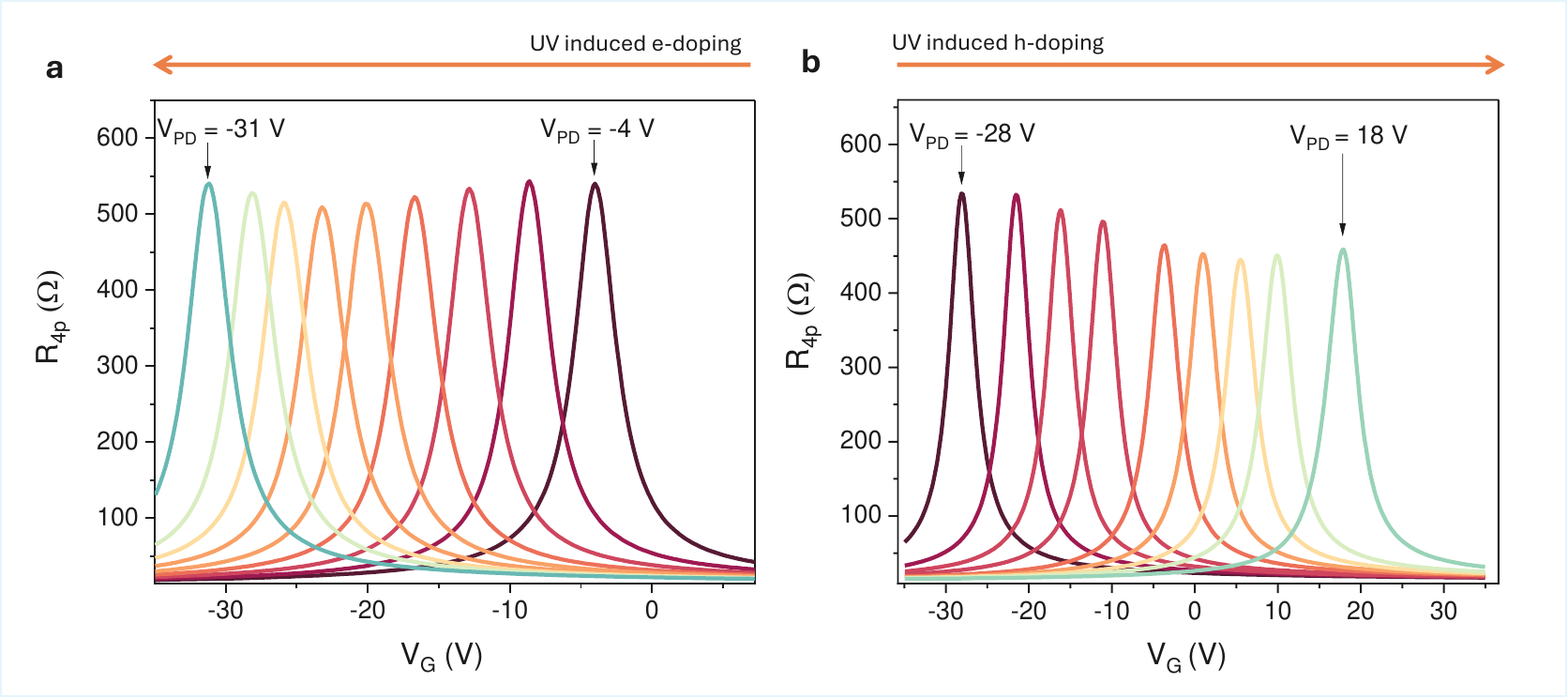}
    \caption{\textbf{Photodoping of electron and holes in graphene using our technique:} (a) $R_\mathrm{4p}-V_\mathrm{G}$ curves of device D3 subjected to UV irradiation at various $V_\mathrm{PD}$. Each transfer characteristic plot corresponds to incremental doping of the graphene channel achieved by increasing the $V_\mathrm{{PD}}$ with $V_\mathrm{PD}=-31$~V being the leftmost and $V_\mathrm{PD}=-4$~V being the rightmost curve. The shift in CNP towards the left-hand side implies photodoping of electrons in the graphene channel. (b) $R_\mathrm{4p}-V_\mathrm{G}$ curves for the same device demonstrating the feasibility of UV-induced hole doping in the channel.}
    \label{fig:electron_hole}
\end{figure}

Just as we have demonstrated UV-induced electron doping of the graphene channel in previous section (indicated in Fig.~\ref{fig:electron_hole}a), we have also shown in Fig.~\ref{fig:electron_hole}b the ability to achieve p-type doping by applying positive back gate voltages ($V_\mathrm{PD}>0$~V), which shifts the CNP towards positive $V_\mathrm{G}$. This behaviour indicates the presence of acceptor-like defect states within the hBN flake, and hBN/SiO$_2$ interfaces capable of transferring holes to the graphene channel. Notably, the doping kinetics reveal a striking asymmetry: n-type doping occurs rapidly (typically within tens of seconds), while p-type doping requires several hundred seconds to achieve comparable CNP shifts. This temporal asymmetry may be attributed to a significantly lower concentration of acceptor-like defect states in hBN, consistent with previous experimental observations \cite{ju2014photoinduced}.

\section{Timescales for different electronic scattering processes in monolayer graphene:}

To investigate the influence of UV-induced disorder on carrier scattering rate in the graphene channel, we analyse the experimental observations obtained from electrical transport primarily using three scattering mechanisms:

\textbf{(i) Short-ranged disorder-mediated scattering}, arising from point-like defects or UV-induced local potential fluctuations within the hBN flakes or across vdW interfaces. This is modelled using a delta potential $U_\mathrm{0}(r)$ $=V_\mathrm{0}\delta (r)$, where $\mathrm{V_0}$ is the strength of delta potential

\begin{equation} \label{eq:short-range scattering}
\tau_\mathrm{s}^{-1}=\frac{1}{4}\frac{n_\mathrm{i}k_\mathrm{F} V_\mathrm{0}^2 }{v_\mathrm{F}}
\end{equation} 

where  $\tau^{-1}_\mathrm{s}$ is the relevant scattering rate, $k_\mathrm{F}$ $=\sqrt{\pi n}$ is the Fermi wavevector, and $v_\mathrm{F}$ is the Fermi velocity. This gives us one possible scaling - $\tau_\mathrm{s}^{-1} \propto \sqrt{n}$ \cite{sarkar2015role}.

\textbf{(ii) Long-ranged disorder-mediated scattering}, e.g., from charged impurities, mainly near the Gr/hBN and hBN/SiO$_2$ interfaces. Due to the action of the Coulomb potential $U_\mathrm{s}(r)$=$eQ/4\pi \epsilon_\mathrm{0} \epsilon_\mathrm{r}r$, the relevant scattering rate ($\tau^{-1}_\mathrm{c}$) is given by  
\begin{equation} \label{eq:long-range scattering}
\tau_\mathrm{c}^{-1}=\frac{u_\mathrm{0}^2}{v_\mathrm{F} k_\mathrm{F}}
\end{equation}

where $u_{0}^2$=$n_\mathrm{i} Q^2 e^2/16 \epsilon_\mathrm{0}^2\epsilon_\mathrm{r}^2 r^2$, Q is the total impurity charge, $\epsilon_\mathrm{r}$ is the relative electric permittivity, and $n_\mathrm{i}$ is the intrinsic charged defect density in the system. The resultant $\tau^{-1}_\mathrm{c}\propto 1/\sqrt{n}$ is a signature of long-ranged impurity-mediated scattering \cite{sarkar2015role}.

\textbf{(iii) Electron-acoustic phonon scattering }, which dominates at lower energies (particularly, the energy range accessible using gate electric fields) at room temperature. The scattering rate ($\tau_\mathrm{{el-ph}}^{-1}$) is given by 

\begin{equation} \label{eq:electron phonon scattering}
\tau_\mathrm{el-ph}^{-1}=\frac{1}{\hbar^3}\frac{E_\mathrm{F}}{4 v_\mathrm{F}^2}\frac{D^2}{\rho_\mathrm{m} v_\mathrm{ph}^2} k_\mathrm{B}T
\end{equation}

where $\rho_\mathrm{m}$ is the mass density of graphene, $v_\mathrm{{ph}}$ is the acoustic phonon velocity, $E_\mathrm{F}$ is Fermi Energy, and ${D}$ is the deformation potential \cite{hwang2008acoustic}. This also yields a familar $n$-dependence: $\tau_\mathrm{el-ph}^{-1}\propto\sqrt{n}$.

\begin{figure}[tbh!]
    \centering
    \hspace*{-1cm}
    \includegraphics[scale=0.8]{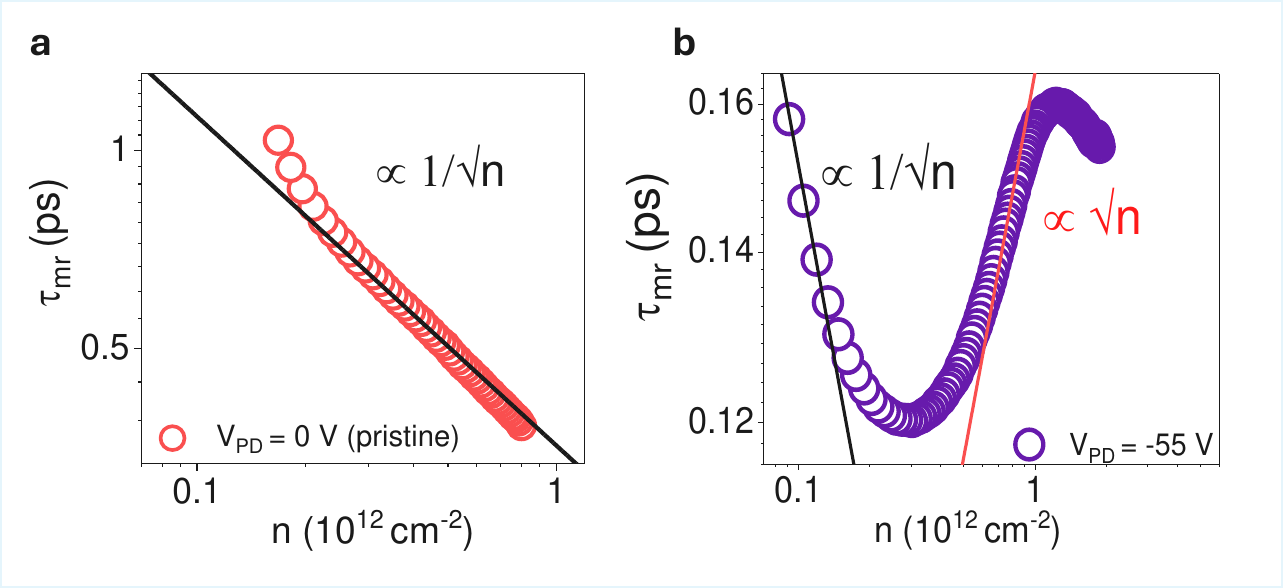}
    \caption{\textbf{Dominant scattering mechanisms in the hBN-encapsulated graphene channel:} (a) $\tau_\mathrm{mr}$ vs n plot for the pristine state ($V_\mathrm{PD}=0$~V) of device D3. The solid black line fits the data, as per Eqn.~\ref{eq:electron phonon scattering} and scales as $\tau_\mathrm{mr} \propto 1/\sqrt{n}$, as expected from electron-acoustic phonon scattering. (b) Variation of $\tau_\mathrm{mr}$ with $n$ at  $V_\mathrm{{PD}}=-55$~V for the same device, where residual charged impurity scattering is considered in addition to the short-ranged disorder mediated scattering and electrons-acoustic phonon scattering to fit the data. Solid black line accounts for the short–ranged scattering given by Eqn.~\ref{eq:short-range scattering}, given by $\tau_\mathrm{mr} \propto 1/\sqrt{n}$ while the solid red line correspond to the additional long-range Coulomb scattering and scales as $\tau_\mathrm{mr} \propto \sqrt{n}$ respectively.} 
    \label{fig:fig_fitting_sup}
\end{figure}

Fig.~\ref{fig:fig_fitting_sup}a shows the experimentally obtained momentum relaxation time $\tau_\mathrm{{mr}}$ as a function of carrier density ${n}$ for the pristine state of the sample. In this regime, electron-acoustic phonon scattering, given by Eqn.~\eqref{eq:electron phonon scattering}, is the dominant scattering mechanism. The extracted value of ${D=15 \pm 0.8}$~eV, from the fitting of Eqn.~\eqref{eq:electron phonon scattering} was found to lie within the range of $10-30$~eV, reported in previous experiments \cite{efetov2010controlling, chen2008intrinsic, mckitterick2016electron}. Fig.~\ref{fig:fig_fitting_sup}b presents a similar analysis for the case when the sample was subjected to UV irradiation at $V_\mathrm{{PD}}=-55$~V. The experimental data exhibits a non-monotonic dependence of $\tau_\mathrm{mr}$ on $n$, suggesting the presence of multiple scattering mechanisms across the entire experimentally accessible range of carrier densities. The UV-irradiated devices harbor activated defect states and charged traps within the hBN flake, along with interfacial disorder across the hBN/SiO$_2$ boundary. The net scattering process involves a combination of short-ranged and long-ranged impurity scattering mechanisms. The solid black line in Fig.~\ref{fig:fig_fitting_sup}b represents the short-ranged impurity-mediated scattering contribution at lower carrier densities, aptly described by Eqn.~(\ref{eq:short-range scattering}), which leads to $\tau_\mathrm{mr}$ scaling as $\propto1/\sqrt{n}$. Further, at higher densities, $\tau_\mathrm{mr}$ starts increasing with increasing $n$. This dependence can be explained using a long-ranged impurity-mediated scattering process - the solid red line shows the best fit, as per Eqn.~(\ref{eq:long-range scattering}), from which we evaluate the value of  $n_\mathrm{i}$ to be $\simeq 0.3\times 10^{12}$~cm$^{-2}$. This value of the fit parameter is in close agreement with other experimentally obtained electrical transport data associated with the present work. 
 
\section{Circuitry for electron noise thermometry}

\begin{figure}[tbh!]
    \centering
    \hspace*{-0.2cm} 
    \includegraphics[scale=0.5]{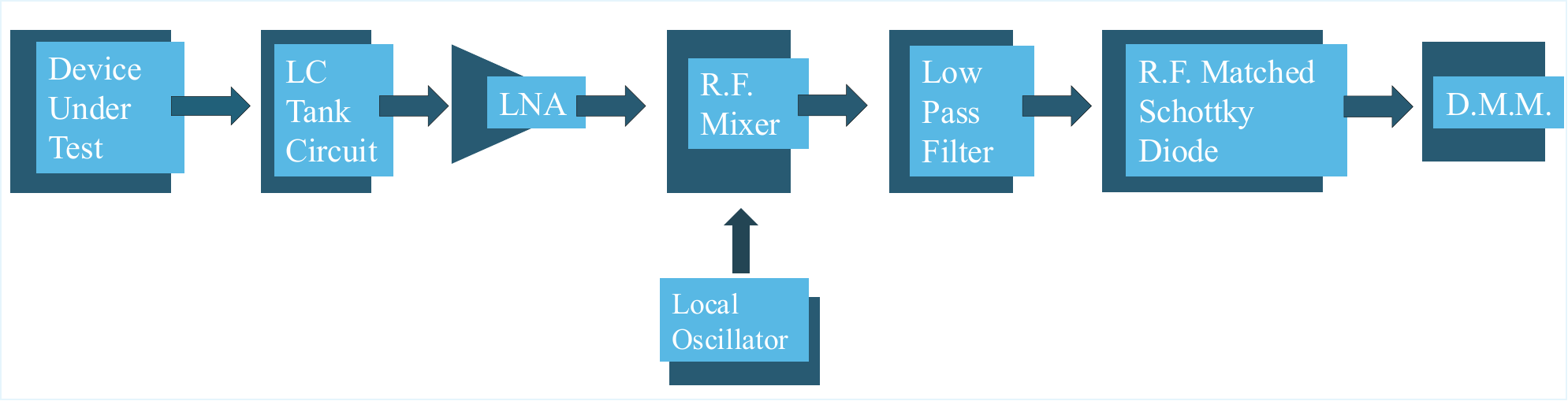}
    \caption{\textbf{Circuit to measure Johnson-Nyquist noise }: The circuit consists of an LC tank circuit coupled to a low-noise amplifier (LNA) to amplify thermal fluctuations from the DUT. A local oscillator and RF mixer down-converts the signal, which is then filtered through a low-pass filter. The Schottky diode, impedance-matched to the RF stage, acts as an envelope detector, rectifying and integrating the noise into a DC voltage read by the digital multimeter (DMM). } 
    \label{fig1}
\end{figure}

We performed thermal noise measurements using a custom-built Johnson-Nyquist noise measurement setup, as shown in Fig.~\ref{fig1}, where the graphene device, referred to as Device Under Test (DUT), was wire-bonded to a printed circuit board incorporating an impedance-matched LC tank circuit.

The RMS voltage associated with the thermal noise produced by the DUT can be expressed as
\begin{equation}
\langle V^2 \rangle = 4k_\mathrm{B} TR \Delta f 
\end{equation}

where $R$ is the two-probe resistance of the DUT and $\Delta f$ is the measurement bandwidth.

The selected L and C, consisting of an air-core inductor $L = 220$~nH and a ceramic parallel-plate capacitor $C =3.9$~pF, yields a resonant frequency given by
\begin{equation}
f_\mathrm{0}=\frac{1}{2\pi}{\sqrt{\frac{1}{LC}-\frac{1}{R^2C^2}}}
\end{equation}

For our circuit, $f_\mathrm{0} \simeq 100$~MHz. This configuration maximises signal transmission across a small frequency band around $f=f_\mathrm{0}$ by ensuring optimal impedance matching at $Z = 50$~$\Omega$ and suppressing other frequency components. The output noise power, after passing the impedance-matching $LC$ circuit, is given by

\begin{equation}
\langle V^2 \rangle = 4k_\mathrm{B} TR \Delta f \times (1-\Gamma^2)
\end{equation}

where $\Gamma$ is the reflection coefficient, given by

\begin{equation}
\Gamma = \frac{Z(\omega) - Z_0}{Z(\omega) + Z_0}
\end{equation}

with $Z(\omega)$ being the device impedance and $Z_0$ is the characteristic impedance of the transmission line. 

The impedance-matched RF signal, which carries the thermal noise of the device, is routed out of the cryostat through RF-compatible (RG-316) coaxial cables to a set of low-noise amplifiers (ZFL-500LN+ and MITEQ-AU 1291) which amplify the noise signal without adding too much external noise, quantified by the noise temperature $T_\mathrm{N}$ of the amplifiers. The amplified signal is subsequently fed into a RF mixer (ZLW-3+) driven by a local oscillator (Keysight N5173B signal generator) with an output signal with frequency $f_\mathrm{0}$. This acts as an analog up- and down-converting device and branches the signal into two frequency components, namely $f+f_\mathrm{0}$ and $f-f_\mathrm{0}$. A low-pass filter (SLP-1.9+) is used to retain the downconverted component of the signal while suppressing the higher-frequency component. The resulting RMS noise voltage is given by

\begin{equation}
\langle V^2 \rangle = 4k_\mathrm{B}GS\Delta f (1 - \Gamma^2)RT
\end{equation}

where $G$ is the amplifier gain and $S$ accounts for the signal loss from the mixer and filter. The filtered signal is then passed through an RF-matched Schottky diode (DMS-104P), which integrates the signal over a specific frequency range and converts the RF power into a DC voltage. The DC voltage $V_\mathrm{DC}$ is finally recorded with a digital multimeter (DMM).  Using information obtained from the diode calibration, the expression for the obtained DC voltage can be written as

\begin{equation}
V_\mathrm{DC}=\alpha\frac{\langle V^2 \rangle}{R}
\end{equation}

where $\alpha$ is the power-to-voltage conversion factor. This leads to

\begin{equation}
V_\mathrm{DC} = 4k_\mathrm{B} \alpha G S \Delta f (1 - \Gamma^2) \dfrac{\sum\limits_{i} T_i R_i}{\sum\limits_{i} R_i}
\end{equation}

since we can write $R = R_\mathrm{ch} + 2R_\mathrm{c}$ as the total two-probe resistance where $R_\mathrm{ch}$ is the graphene channel resistance with $T_\mathrm{ch}$ as the electronic temperature of the graphene channel and $2R_\mathrm{c}$ as the contact resistance with $T_\mathrm{c}$ is the temperature of the gold electrical contacts. This finally leads to

\begin{equation} \label{lastjneqn}
V_\mathrm{DC} = 4\alpha  G S k_\mathrm{B} \Delta f \left(1 - \Gamma^2 \right) \frac{R_\mathrm{ch}T_\mathrm{ch} + 2R_\mathrm{c}T_\mathrm{c}}{R_\mathrm{ch} + 2R_\mathrm{c}}
\end{equation}

Using Eqn.~\ref{lastjneqn}, we get the value of the electronic temperature $T_\mathrm{ch}$ of the graphene channel.

Further, we adopt the cold phonon bath approximation, wherein Joule heating of the graphene channel under DC bias is assumed to be ineffective in changing the temperature of the contacts, owing to the large thermal boundary resistance at the metal-graphene interfaces \cite{roukes1985hot} and high thermal conductivity of gold - implying that the electrical contacts stay at room temperature ($T_\mathrm{c}=296$~K). Other factors in the expressions such as $\alpha$, $G$, $S$, $\Delta f$ are determined from the calibration procedure and known experimental settings.

The sensitivity and resolution of this technique are fundamentally limited by Dicke's radiometer formula. It sets a bound on the temperature resolution ($\delta T$) of such a noise measurement and is given by:

\begin{equation}
\delta T = \frac{T_\mathrm{sample}+T_\mathrm{N}}{\sqrt{\tau \Delta f}}
\end{equation}

where $T$ is the lattice temperature of the device ($=300$~K), $T_\mathrm{N}$ is the system noise temperature (primarily from the amplifier =$75$~K ), $\Delta f$ is the measurement bandwidth ($\simeq 4$~MHz), and $\tau$ is the integration time ($=1.5$~s). This gives $\delta T = 150$~mK at room temperature and $\delta T = 5$~mK at $10$~K. To experimentally measure the error associated with our readings, we record around $1000$ data points corresponding to a fixed value of the number density and dc electric field across the channel. The standard deviation calculated from these points helps us to calculate the error in $\mathcal{L}$ and $\mathcal{\sigma}_\mathrm{Q}$.

\section{Discrepancy in effective Lorentz ratio between forward and backward $V_\mathrm{G}$ sweeps:}

\begin{figure}[tbh!]
    \centering
    \hspace*{-1.0cm} 
    \includegraphics[scale=0.7]{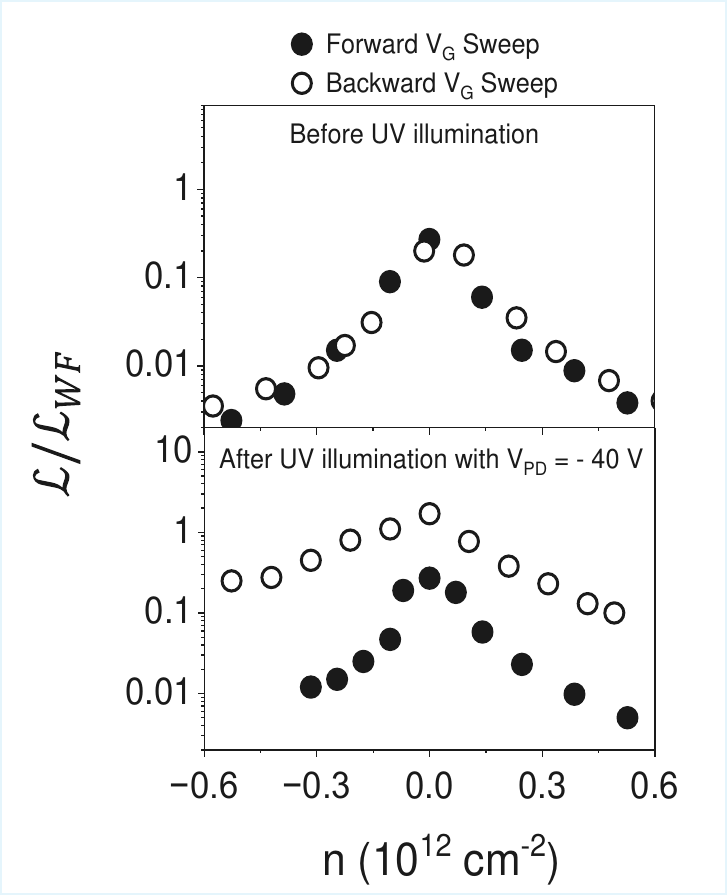}
    \caption{\textbf{Effect of UV illumination on Lorentz ratio between forward and backward $V_\mathrm{G}$ sweeps:} Variation of $\mathcal{L}/\mathcal{L}_\mathrm{WF}$ as a function of $n$ in pristine condition (Top panel) and at $V_\mathrm{{PD}} = -40$~V (Bottom panel) for device D2 at $296$~K. Data for both forward (indicated by solid circles) and backward $V_\mathrm{G}$ sweeps (indicated by hollow circles) have been shown for each dataset.} 
    \label{fig7}
\end{figure}

On application of UV radiation, thermal transport in the backwards $V_\mathrm{G}$ sweeps gets enhanced by additional trap- and disorder-assisted mechanisms \cite{chen2008intrinsic, cadore2016thermally}. These arise due to the ionisation of defect states in the hBN layers and at the Gr/hBN and hBN/SiO$_2$ interface during UV exposure, under a non-zero photodoping voltage ($V_\mathrm{PD} \ne 0$~V). The resulting disorder leads to asymmetric scattering environments for the forward and backwards $V_\mathrm{G}$ sweeps. This discrepancy in $\mathcal{L}/\mathcal{L}_\mathrm{WF}$ vs n between the forward and backward $V_\mathrm{G}$ sweeps (Fig.~\ref{fig7}: Bottom Panel) increases with increasing $V_\mathrm{PD}$. At carrier densities away from the charge neutrality point, the backwards $V_\mathrm{G}$ sweeps exhibit an enhanced electronic thermal conductivity $\kappa_\mathrm{e}$, which becomes significantly larger than that in the forward $V_\mathrm{G}$ sweep.

While enhanced electron-impurity scattering reduces both electrical ($\sigma$) and thermal ($\kappa_\mathrm{e}$) conductivities, their distinct transport mechanisms yield different responses. Electrical conductivity depends on directed electron flow, where frequent scattering events strongly disrupt net charge transport, leading to significant $\sigma$ suppression. In contrast, thermal conductivity relies on local energy transfer through electronic collisions, which persist even with increased scattering, resulting in a more moderate $\kappa_\mathrm{e}$ reduction. This fundamental difference explains the observed rise in $\mathcal{L}/\mathcal{L}_\mathrm{WF} = \kappa_\mathrm{e}/\sigma T$ during backward $V_\mathrm{G}$ sweeps- as disorder amplifies, $\sigma$ decreases more sharply than $\kappa_\mathrm{e}$, driving the ratio toward the Wiedemann-Franz limit ($\mathcal{L}/\mathcal{L}_\mathrm{WF}$ $\to$ $1$) of unity. The effect is particularly pronounced in backwards $V_\mathrm{G}$ sweeps where activated interface traps maximize scattering asymmetry. In contrast, the forward $V_\mathrm{G}$ sweep continues to reflect relatively pristine transport characteristics, where the system remains more ballistic or hydrodynamic in nature, thereby deviating from the WF limit. This directional dependence highlights that the addition of disorder to a clean graphene channel primarily influences the thermal transport during the backwards $V_\mathrm{G}$ sweep, while leaving forward $V_\mathrm{G}$ sweep characteristics relatively unaffected. 

\section{Contribution of holes to the electrical transport in UV-doped graphene}

\begin{figure}[tbh!]
    \centering
    \hspace*{-0.2cm}
    \vspace{0cm}
    \includegraphics[scale=0.7]{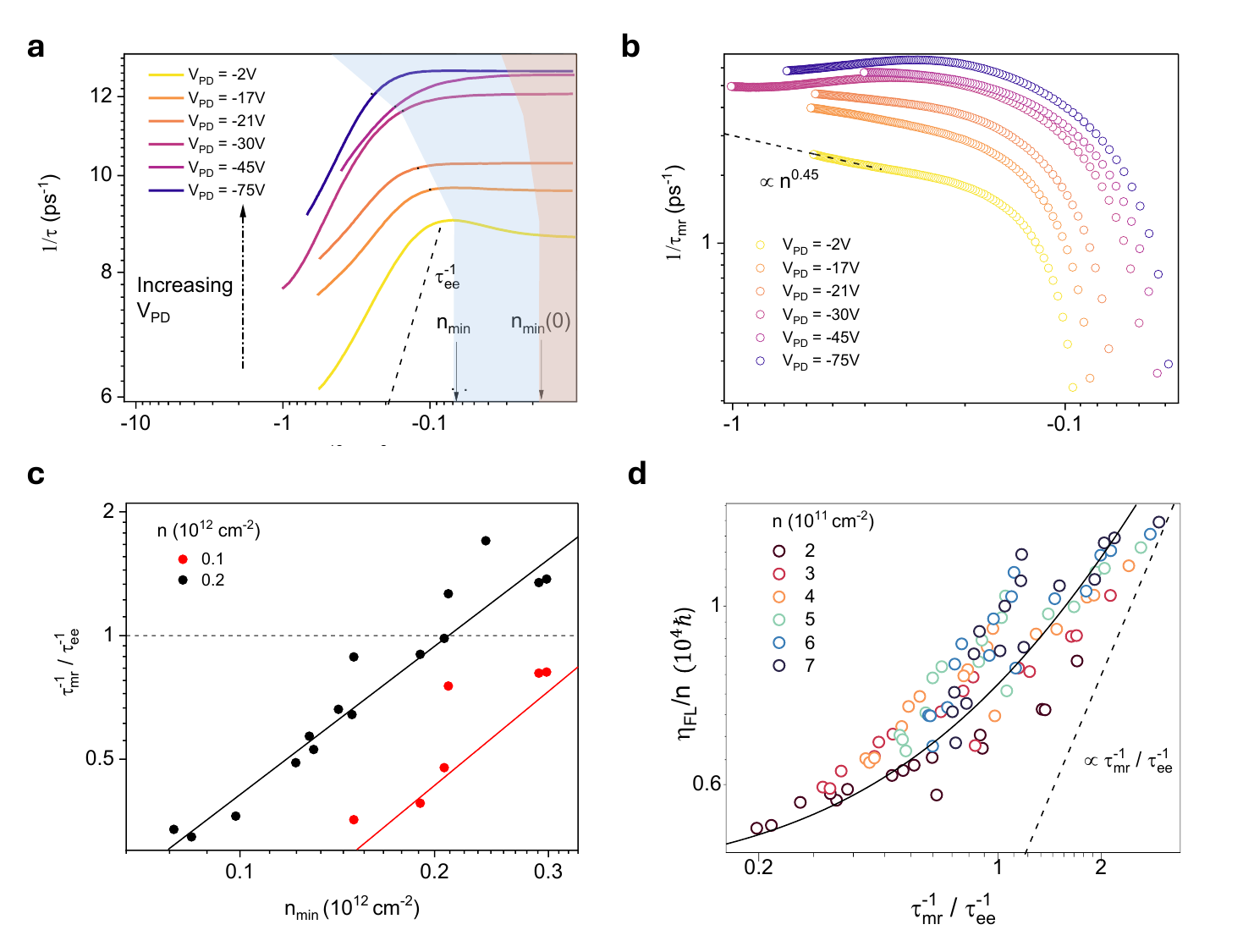}
    \caption{\textbf{Scaling trends observed in the hole doped regime:} (a) Variation of the total scattering rate $\tau^{-1}$ as a function of $n$ for different values of $V_\mathrm{PD}$. The dashed line shows the variation of electron-electron relaxation scattering rate $\tau_\mathrm{ee}^{-1}$ with $n$. The $n<n_\mathrm{min}(0)$ region is colored with light purple, whereas $n_\mathrm{min} (0) \le n \le n_\mathrm{min}$ region is colored with light blue. (b) Variation of the momentum relaxation scattering rate $\tau_\mathrm{mr}^{-1}$ as a function of $n$ for the same values of $V_\mathrm{PD}$, as those depicted in panel (a). The solid line indicates a $\sim \sqrt{n}$ dependence. (c) Ratio of momentum relaxation scattering rate ($\tau_\mathrm{mr}^{-1}$) to the electron-electron scattering rate ($\tau_\mathrm{ee}^{-1}$) as a function of $n_\mathrm{min}$ for two different carrier densities. The solid lines vary as $n_\mathrm{min}$ and serve as a guide to the eye. The region corresponding to $\tau_\mathrm{mr}^{-1}/\tau_\mathrm{ee}^{-1} < 1$ indicates the viscous limit, while the region with $\tau_\mathrm{mr}^{-1}/\tau_\mathrm{ee}^{-1} > 1$ indicates the diffusive limit. (d) $\eta_\mathrm{FL}/n$ as a function of $\tau_\mathrm{mr}^{-1}/\tau_\mathrm{ee}^{-1}$ for different carrier densities, indicated using symbols of different color. The solid black line highlights asymptotic behaviour of $\eta_\mathrm{FL}/n$ in the viscous regime and $\propto \tau_\mathrm{mr}^{-1}/\tau_\mathrm{ee}^{-1}$ in the diffusive regime. The dotted line indicates $\eta_\mathrm{FL}/n \propto \tau_\mathrm{mr}^{-1}/\tau_\mathrm{ee}^{-1}$ in the diffusive limit,  and serves as a guide to the eye.   }
    \label{R2Q5}
\end{figure}

In the low-density regime, the electron-hole plasma dominates the scattering process, whereas transport in the high-density regime is governed by the majority carriers, either electrons or holes. In the main file, we have presented data and results for quasiparticle scattering in both the low and high carrier density regimes. However, for discussing the physics at higher densities, we have limited ourselves only to the electron-doped regime, since the hole-doped regime leads to the formation of p-n junctions across the metal-graphene interface, leading to the suppression of momentum-conserving scattering and thereby yielding slightly inconsistent results. Nevertheless, the extremely clean nature of our samples allowed us to perform a similar analysis to understand the behaviour of holes, as indicated in Fig.~\ref{R2Q5} - almost all of the results that were observed to be true for the electron-doped regime were found to be in good agreement with those observed in the hole-doped regime as well. This indicates that disorder induced by UV-doping affects both hole-doped and electron-doped conduction equally in hydrodynamic graphene.
 
\section{Carrier density dependence of shear viscosity and kinematic viscosity in pristine graphene:}

\begin{figure}[tbh!]
    \centering
    \hspace*{-1.0cm} 
    \includegraphics[scale=0.75]{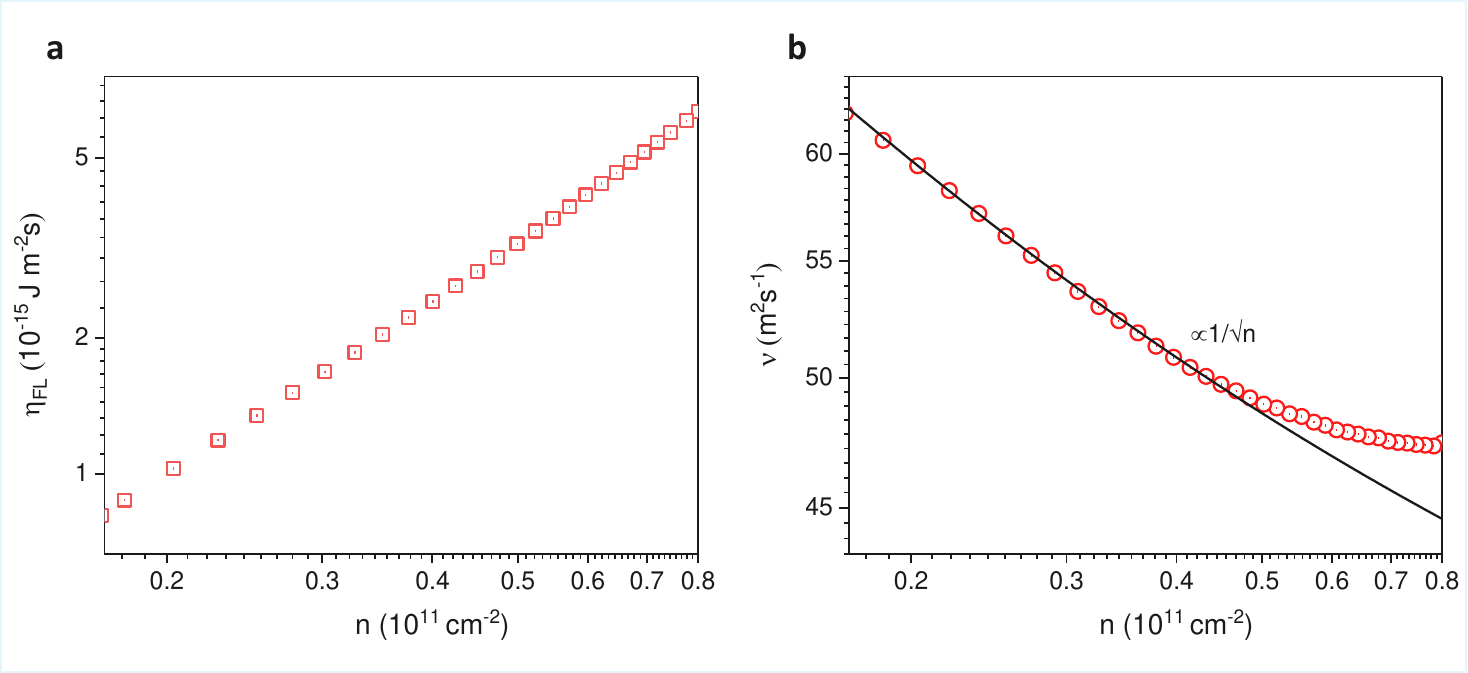}
    \caption{\textbf{Variation of shear viscosity and kinematic viscosity as a function of carrier density:} (a) Variation of $\eta_\mathrm{FL}$ as a function of n for the pristine state of device D3 at ${296}$~K. (b) ${\nu}$ as a function of $\mathrm{n}$ for the same device. The black line serves as a guide to the eye, indicating ${1/\sqrt{n}}$ dependence. } 
    \label{fig8}
\end{figure}

The shear viscosity of the pristine monolayer graphene channel was calculated using the expression $\eta_\mathrm{FL} = n^2e^2W^2/12\sigma$, as shown in Fig.~\ref{fig8}a, where $n$ is the carrier density, e is the elementary charge, $\sigma$ is the electrical conductivity, and W is the width of the graphene channel. Subsequently, the kinematic viscosity ($\nu$) for the pristine graphene channel has been extracted from Fig.~\ref{fig8}a, using the expression $\nu=\eta_\mathrm{FL}/{mn}$, where $m$ is the effective mass of carriers. The experimentally obtained results (indicated in Fig.~\ref{fig8}) for the pristine graphene show a $1/\sqrt{n}$ dependence, whereas Fermi liquid theory predicts ${\nu \propto \sqrt{n}}$ dependence. However, this observation falls in line with previous reports, which have recorded a monotonically decreasing trend in $\eta_\mathrm{FL}$ or $\nu$ as a function of $n$ \cite{talanov2024observation, bandurin2016negative}.

\bibliography{ref}

\newpage

